\documentclass[12pt]{iopart}
\expandafter\let\csname equation*\endcsname\relax 
\expandafter\let\csname endequation*\endcsname\relax

\usepackage[dvipsnames]{xcolor}
\usepackage{
 amsmath,
 soul,
 amsfonts,
 color,
 graphics,
 hyperref, 
 textgreek, 
 }
\usepackage[utf8]{inputenc}

\usepackage{todonotes}
\presetkeys{todonotes}{size=\footnotesize, color=yellow!90}{}

\newcommand{\ket}[1]{\left|#1\right>}

\newcommand{\omegarf}{\omega_\text{rf}}
\newcommand{\Omegarf}{\Omega_\text{rf}}
\newcommand{\OmegaL}{\Omega_\text{L}}

\renewcommand{\ll}{\langle\langle}

\begin{document}

\title{Matter-wave interferometers using TAAP rings}

\author{P. Navez$^1$, S. Pandey$^{2,3}$, H. Mas$^{2,4}$, K. Poulios$^2$, T. Fernholz$^5$, and W. von Klitzing$^2$ }

\address{
$^1$ Crete Center for Quantum Complexity and Nanotechnology, Depart. of Physics, University of Crete, Heraklion 70113, Greece\\
$^2$ Institute of Electronic Structure and Laser, Foundation for Research and Technology - Hellas, Heraklion 70013, Greece\\
$^3$ Depart. of Materials, Science and Technology, University of Crete, Heraklion 70113, Greece\\
$^4$ Depart. of Physics, University of Crete, Heraklion 70113, Greece\\
$^5$ School of Physics and Astronomy, University of Nottingham, Nottingham NG7~2RD, United Kingdom}
\ead{TAAPs@bec.gr}

\begin{abstract}
We present two novel matter-wave Sagnac interferometers based on ring-shaped time-averaged adiabatic potentials (TAAP). 
For both the atoms are put into a superposition of two different spin states and manipulated independently using elliptically polarized rf-fields. 
In the first interferometer the atoms are accelerated by spin-state-dependent forces and then travel around the ring in a  matter-wave guide.
In the second one the atoms are fully trapped during the entire interferometric sequence and are moved around the ring in two spin-state-dependent `buckets'.
Corrections to the ideal Sagnac phase are  investigated for both cases. 
We experimentally demonstrate the key atom-optical elements of the interferometer  such as the independent manipulation of two different spin states in the ring-shaped potentials under identical experimental conditions.
\end{abstract}


\maketitle
\newpage
\section{Introduction} \label{sec:intro}

\subsection{Interferometry with Atomic clocks}

Since its first demonstration about 20 years ago \cite{ Carnal1991PRL, Keith1991PRL,   Kasevich1991PRL, Barrett2014CRP} atom interferometry is one of the most sensitive and accurate forms of inertial sensing.
It carries the promise of greatly enhanced sensitivity and precision, which could be exploited for fundamental physics \cite{Will2006LRIR}, geo-sensing \cite{Igel2007GJI} and inertial navigation \cite{Tino2014BOOK}.
Most atom interferometers use free-space atomic beams \cite{Berg2015PRL,Barrett2014CRP} with Raman or Bragg beam splitters.
The sensitivity of the Sagnac interferometer scales with the enclosed area, which is limited by the momentum difference between the two arms and the available time-of-flight of the free-falling atoms.  
Guiding the atoms through an atomtronic circuit can provide very large areas and thus pave the way towards compact, ultra-sensitive matter-wave interferometers.
A number of such circuits have been proposed and some demonstrated, e.g.~using dipole potentials~\cite{Bongs2001CRDLADSSI,Henderson2009NJOP,Heathcote2008NJOP}, magnetic traps on atom-chips~\cite{Hansel2001N,Schumm2005NP,Dumke2002PRL}, and using time-averaged adiabatic potentials (TAAPs) \cite{Lesanovsky2007PRL,Sherlock2011PRA,Gildemeister2010PRA, Gildemeister2012PRA}.
Michelson type interferometers have also been demonstrated over  distances of tens of micrometers~\cite{Schumm2005NP,Wang2005PRL,Berrada2013NC,Schumm2011IWBCOAC}.
However, an experimental demonstration of coherent atom guiding over macroscopic distances with interferometric stability has proven to be an elusive  goal.
The main reason has been the corrugation of the guiding potentials~\cite{Henkel2003APB,Trebbia2007PRL, Sinclair2005PRA}.

Another challenge for atomtronic circuits is the coherent splitting of the atomic cloud inside the waveguide. 
In the case of the standard Bragg beam splitter, the reference with respect to which the interferometer measures the rotation, is set by the phase difference of the optical beams.
Unfortunately, a drift of the atomic waveguide  relative to the Bragg beams can result in a phase shift at the readout of the interferometer, which would be interpreted as an acceleration.
Coherent splitting of Bose-Einstein condensates has previously been demonstrated on atomchips by transforming a magnetic trap into a double-well in the direction of tight confinement \cite{Schumm2005NP}. 
Since the splitting process has to be very slow compared to the trapping frequency in the direction of the splitting, this  method  would be prohibitively slow in the longitudinal direction.
Similarly, the state-dependent manipulation of atoms can be achieved  using direct microwave dressing on magnetic atom-chips  \cite{Bohi2009NP,Guarrera2015NJOP}.
Recently, a clock-type interferometer has been proposed, based on state-dependent manipulation of atoms on an atom-chip~\cite{Stevenson2015PRL, Fernholz2007PRA}.  A superposition of two atomic hyperfine (clock) states is created using an rf/microwave pulse. The hyperfine components are then moved separately in opposite directions around an enclosed area using rf fields. They are finally recombined using a second rf/microwave pulse, with the interferometric signal being the population difference between these hyperfine states.
In order to realize a full loop Sagnac interferometer in this way \cite{Stevenson2015PRL} a ring-shaped structure on an atomic micro-chip can be used, which is dressed using strong rf-fields with a doughnut shaped polarisation structure~\cite{Fernholz2007PRA}.

In the present paper, we study a guided clock-type interferometer based on ultra-cold thermal atoms or Bose-Einstein Condensates (BECs) in time-averaged adiabatic potentials (TAAPs) \cite{Lesanovsky2007PRL}. By clock-type interferometer we mean an interferometer, which utilizes two distinct states having different eigenenergies, 
These potentials have the advantage of being extremely smooth, controllable and flexible. 
We will examine two types of interfero\-meters, one where thermal or Bose-condensed atoms propagate freely along a ring-shaped TAAP waveguide and a second where the atoms are fully confined and moved along a circular path.

The paper is structured in the following way: Section\,\ref{sec:TrappingAndGuidingPotentials} reviews time averaged potentials and extends them to arbitrary polarization states for interferometry. In section\,\ref{sec:TAAP_Interferometers} the Sagnac phase is briefly introduced and the different interferometric schemes are presented. TAAP interferometers based on Bragg beams splitters are discussed in section\,\ref{sec:BraggRingInterferometer} while section\,\ref{sec:StateDependentInterferometer} deals with TAAP interferometers based on rf/microwave beam splitters with subsequent state-dependent acceleration (section\,\ref{sec:StaticAccelerator}) or use of moving atom buckets (section\,\ref{sec:movingBucket}). The paper concludes in section\,\ref{sec:Conclusions}.

\begin{figure}
\begin{center}
\includegraphics[width=8cm]{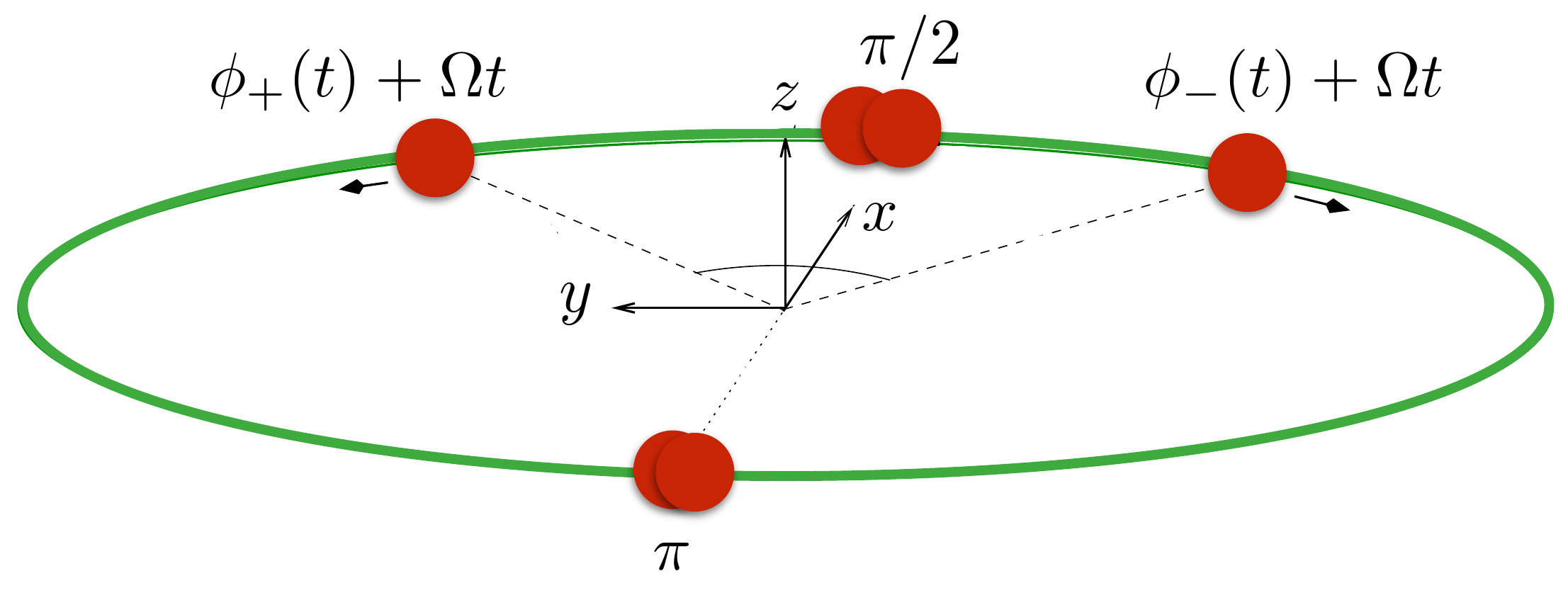}
\caption{Sketch  of the Sagnac interferometer sequence. The $\pi$ and $\pi/2$ refers to the two photon pulses.}
\label{fig:theRing}
\end{center}
\end{figure}

\section{The trapping and guiding potentials}\label{sec:TrappingAndGuidingPotentials}
%
\subsection{Time averaged adiabatic potentials (TAAPs)}\label{sec:TAAPs}
TAAPs  can produce extremely controllable and smooth potentials, making them ideal candidates for matter-wave interferometry.
They have been proposed in \cite{Lesanovsky2007PRL} and realised in \cite{Heathcote2008NJOP, Sherlock2011PRA, Gildemeister2010PRA}.
In the TAAPs used here, a strong rf-field $(B_\text{rf})$ dresses atoms in a static magnetic quadrupole field $(B_\text{Q})$, leading to adiabatic potentials \cite{Zobay2001PRL}.
An oscillating homogeneous magnetic field $(B_\text{m})$ is then added, which effectively leads to a displacement of the quadrupole.
If the oscillation frequency is low compared to the Larmor frequency but fast compared to the center of mass motion of the atoms, then the external potential confining the atoms is the time-average over the modulated adiabatic potentials.
The trapping and guiding structures of TAAPs are generated using large coils located far way from the waveguide itself, thus virtually eliminating any corrugation in the potential.
This makes TAAP rings extremely smooth \footnote{Random corrugations in a potential decay exponentially over the length scale of their characteristic size.}, and therefore ideal candidates for the coherent matter-wave guiding over macroscopic distances and hence for matter-wave interferometry.
 
Let us now consider the Hamiltonian
\begin{equation}
\hat H_\pm= \pm g \mu_\text{B} {\hat {\bf F}}. ({\bf B}({\bf r},t) +{\bf B}_\text{rf}(t))
\,,
\end{equation}
where $\hat {\bf F}$ is the total spin of the  atoms, $\mu_\text{B}$ the Bohr magneton, and $g$ the Land\'e factor. We focus on the $\ket{F,m_\text{F}}=\ket{1,1}$ and $\ket{2,1}$  states of Rubidium 87 and assume that the fields are small enough to neglect the quadratic Zeeman effect. Note that even for larger fields the differential energy shift due to quadratic Zeeman effect can be cancelled \cite{Kazakov2015PRA}.
We combine a static quadrupole field ${\bf B}_\text{Q}({\bf r})=\big(B_\text{x},B_\text{y},B_\text{z}\big)=\alpha\big(x, y, -2 z\big)$ with a homogeneous modulation field which can be tilted by an angle $\delta$ from the vertical (z) axis towards the x axis: ${\bf B}_\text{m}(t)=B_\text{m} \sin(\omega_\text{m} t)\big(\sin(\delta),0,\cos(\delta)\big)$. The total field is then ${\bf B}({\bf r},t)={\bf B}_\text{Q}({\bf r})+{\bf B}_\text{m}(t)$. 
Finally, we add an elliptically polarized rf field ${\bf B}_\text{rf}(t)=\big((B_+ - B_-)\cos(\omegarf t), 0, (B_+ + B_-) \sin(\omegarf t)\big)$. Note that $\omegarf$ itself can be time dependent, i.e.~$\omegarf(t)$, with its frequency $\omegarf \gg \omega_\text{m}$ typically being in the MHz range.
We define the ellipticity parameter  $s=(B_+ - B_-)/(B_+ + B_-)$, where $s=0$ denotes a linear rf in the $z$-axis, and $s=1$ means $B_-=0$ and thus a pure circular rf in the $x-z$ plane. In this paper, the polarization will be almost purely linear with a small admixture of circular rf $(s\,\ll\,1)$. 
The rf field can be understood as the sum of a 
clockwise $(B_{+})$ and an anticlockwise $(B_{-})$ rotating field. 
We can calculate the coupling rates for both the $F=2$ and the $F=1$ hyperfine levels 
by decomposing the rf field into a basis, which is co-moving with the modulation field. 
The resulting dressed states will henceforth be referred to as 
$\ket{2,1}$ and $\ket{1,1}$ states respectively.
In the rotating wave approximation (RWA), we neglect the component of the rf parallel to the B-field and decompose the orthogonal components into two counter-rotating fields, of which we only use the component that adheres to the conservation of angular momentum \footnote{This can be calculated conveniently using the Jones formalism developed for optical polarizations.}. 
The resulting Rabi equation is then:
\begin{equation}\label{eq:rabi}
\begin{aligned}
V_\pm ({\bf r},t)&=&\hbar \sqrt{\left[\OmegaL({\bf r},t)-\omegarf(t)\right]^2+ \Omega_{\pm}^2({\bf r},t)} \\
&\simeq &\hbar\Omega_{\pm}({\bf r},t) +\hbar\frac{\left[\Omega_\text{L}({\bf r},t)-\omegarf(t)\right]^2}{2\Omega_{\pm}({\bf r},t)}\,\,,
\end{aligned}
\end{equation}
where $ \Omega_\text{L}({\bf r},t)=g_\text{F} \mu_\text{B} |{\bf B}|/\hbar$ is the Larmor frequency.
The Rabi coupling (see Appendix \ref{sec:ApppendixAPotential}) is:
\begin{equation}\label{eq:coupling}
\hbar \Omega_{\pm}({\bf r},t)=\hbar \Omega_{0}
\sqrt{\left(1\pm s\frac{B_\text{y}}{|\bf{B}|}\right)^2 - (1-s^2)\frac{B_\text{z}^2}{|\bf{B}|^2}}\,\,,
\end{equation}
where $\hbar\Omega_{0}=\frac{1}{2}g_\text{F}\mu_\text{B} (B_+ + B_-)$, the `$+$' refers to the dressed $\ket{2,1}$ state and the `$-$' to the $\ket{1,1}$ state. $\bf B$, $B_\text{y}$ and $B_\text{z}$ depend both on $\bf r$ and $t$.

\subsection{Experimental Setup} 
\label{sec:expSetup}

The potentials for the ring-TAAPs are formed from a quadrupole field, homogeneous time averaging fields, and rf-dressing fields.
The experiments are performed within a glass vacuum chamber with an estimated pressure in the low $10^{-11}$~Torr range with a vacuum limited life time of more than two minutes.
The quadrupole field of up to 400\,G/cm is generated using a pair of water cooled coils in anti-Helmholtz configuration. 
The modulation field oscillates at 5\,kHz and can have up to 48\,G in amplitude.
It is generated by two pairs of rectangular Helmholtz coils in the horizontal and a pair of round coils in the vertical direction.
The coils are driven from an eight channel 24\,bit sound card, which is amplified using a standard two kilo-watt audio amplifier. 
In order to control the impedance seen by the amplifier we place capacitors and water cooled $2\,\Omega$ resistors in series with the coils.
The three interlocking rf-coils each consist of two single loops in Helmholtz configuration. 
Impedance matching is achieved by placing the pair of coils and the capacitors in parallel. The coils are driven by a 25\,W amplifier (Amplifier Research 25A250A), which is controlled by two arbitrary waveform generators (Tektronix AFG3022).
Both the rf-coils and modulation coils can produce arbitrarily polarized fields in three dimensions. However, for reasons of experimental simplicity, we keep the frequency of the rf-fields constant. The modulation fields can be changed in amplitude and polarization on a time-scale of 3\,ms.

The experimental sequence starts by loading a 3D-MOT with $2\times 10^{9}$  atoms $\left(^{87}Rb\right)$ from a 2D-MOT.
We transfer the atoms to a matched quadrupole time-orbiting potential (TOP). 
We then compress the trap by first increasing the quadrupole gradient and then lowering the amplitude of the modulation field. 
At the same time we cool the atoms using radio-frequency (rf) evaporation to a temperature of about $3T_{c}$ (with $T_{c}$ being the critical temperature).
When the modulation field has reached 5\,G, we switch on the rf-dressing fields at a frequency of ~2.62\,MHz and load the atoms into the dressed potential, which we then deform into the ring-shaped trap.
Finally, we take resonant absorption images of the atomic clouds using a $4f$ custom objective and a CCD camera after a short time of flight.

\subsection{The ring waveguide ($s=0$, $\delta=0$)} \label{sec:pureRingTrap}
\begin{figure}[h!]
\begin{center}
 \includegraphics[width=0.8 \textwidth]{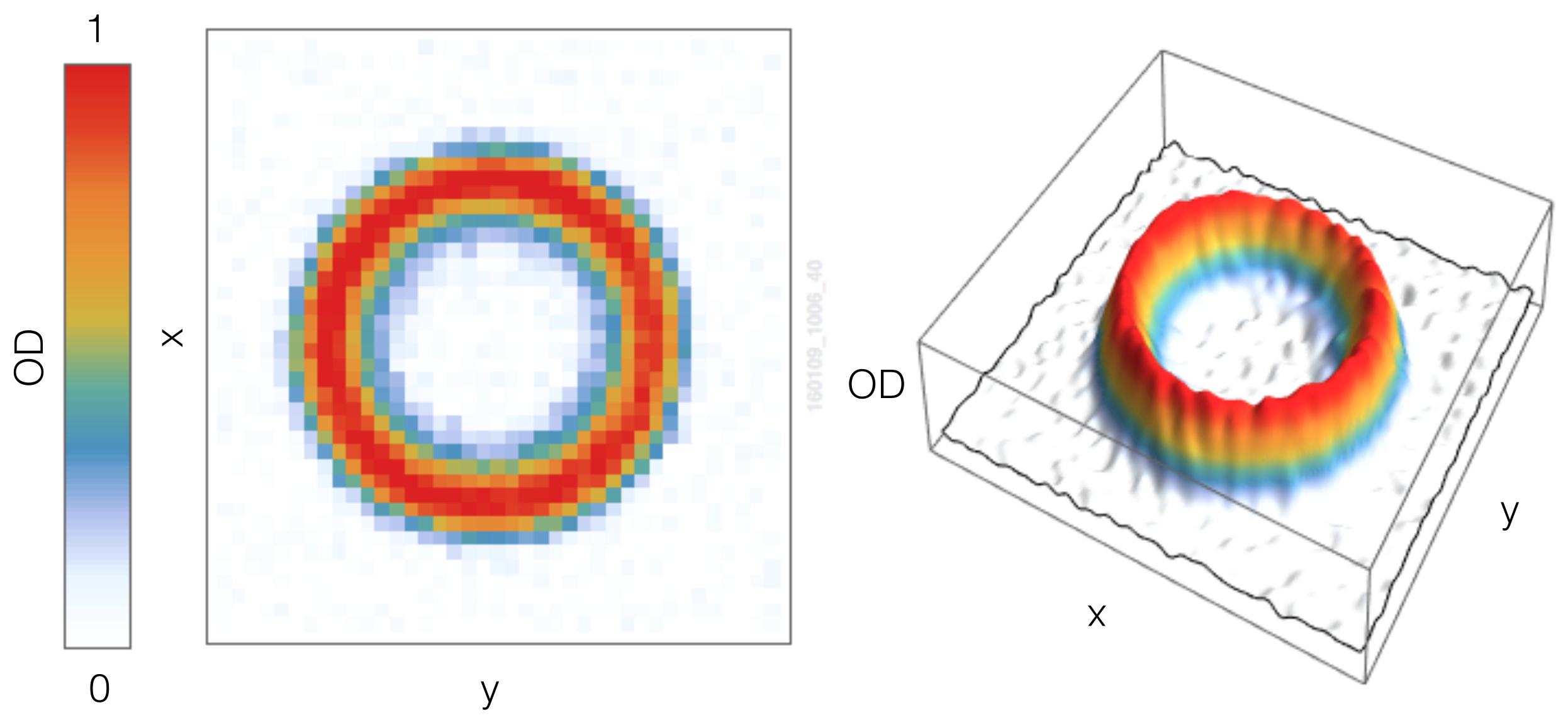}
 \caption{Experimental realisation of a ring-shaped TAAP waveguide $(s=0,\delta=0)$ with Rubidium atoms in the $\ket{2,2}$ state. The quadrupole gradient is $\alpha=$50\,G/cm with $\omegarf/2\pi=2.62$\,MHz. The measured Rabi frequency is $\Omegarf/2\pi=215$\,kHz. The radius of the ring is $R=570$\,\textmu m. 
 }
 \label{fig:ringFlat}
\end{center}
\end{figure}
\noindent In order to trap the atoms  in the vertical direction as well, we add the modulation field ${\bf B}_\text{m}(t)$. 
If the frequency of the modulation field $(\omega_\text{m})$ is fast compared to the trapping frequency but slow compared to the Larmor frequency, then we can time-average the dressed, modulated quadrupole potential:
\begin{equation}\label{eq:VeffIntegral}
V_\pm^\text{eff} ({\bf r})=\frac{\omega_\text{m}}{2\pi}
\int_0^{2\pi/\omega_\text{m}}\!\!\! dt \, V_\pm ({\bf r},t)
\end{equation}
For a purely linearly polarized rf-field ($s=0$) and a purely vertical modulation field
($\delta=0$) the field is cylindrically symmetric with respect to the z-axis and identical for the $\ket{2,1}$ and $\ket{1,1}$ states. 
The potential then forms in the x-y plane a ring-shaped trap of radius $R=\hbar \omega_\text{rf0}/g\mu_B \alpha$.
For mathematical simplicity we modulate the frequency $(\omegarf)$ and amplitude $(B_+(t)=B_-(t))$ of the rf-field such that at ${\bf r}=(R,0,0)$ it stays resonant $(\OmegaL({\bf r},t)-\omegarf(t)=0)$ at a constant Rabi-frequency $(\Omegarf({\bf r},t)=\Omega_{0c}={\text const.})$. 
This can be achieved using the following time dependencies
$\omega_\text{rf}(t)=\omega_\text{rf0}\sqrt{1+\beta^2\sin^2(\omega_m t)}$
and $\Omega_{0}= \Omega_{0c}\sqrt{1+\beta^2\sin^2(\omega_\text{m} t)}$.
In the current experiments constant rf-frequency and amplitude are used.
The ring can then be described close to the trap bottom as~\cite{Lesanovsky2007PRL}:
\begin{equation}\label{eq:ring}
V_\text{ring}^\text{eff} (r,z)=\hbar \Omega_{0c}+\frac{1}{2} m\omega_\text{r}^2(r-R)^2+\frac{1}{2}m\omega_\text{z}^2z^2
\end{equation}
with trapping frequencies in the radial and the z axis directions:
\begin{equation}\label{eq:trappingFrequencies}
\begin{aligned}
\omega_\text{r}&=\omega_0 
(1+\beta^2)^{-1/4}
\\ 
\omega_\text{z}&=2\omega_0 
\left[1-(1+\beta^2)^{-1/2}
\right]^{1/2},
\end{aligned}
\end{equation}
where $\beta=g_\text{F}\mu_\text{B}B_\text{m}/\hbar\omega_\text{rf0}$ is the normalized modulation index and $\omega_0=m_\text{F} g_\text{F}\mu_\text{B}\,\alpha\, \left(m\,\hbar \Omega_\text{0c}  \right)^{-1/2}$.

Figure \ref{fig:ringFlat} shows an absorption image of $4.8\times 10^5$ atoms in a TAAP ring of 1.15\,mm diameter.  Note that the noise in the image is purely due to the imaging system.
 
One of the most useful features of the TAAP rings --- next to their exceptional smoothness --- is the ability to tune the sensitivity of the interferometer simply by changing either the dressing frequency or the gradient of the static quadrupole field $(\alpha)$, which determine the ring's radius and therefore the enclosed area. 
We have demonstrated rings from 400\,\textmu m to 2.6\,mm in diameter filled with up to $6\times 10^5$ atoms at temperatures down to 2\,\textmu K.  
The life time of atoms in such a trap is of the order of ten seconds.

\begin{figure}[h!]
\begin{center}
 \includegraphics[width=9cm]{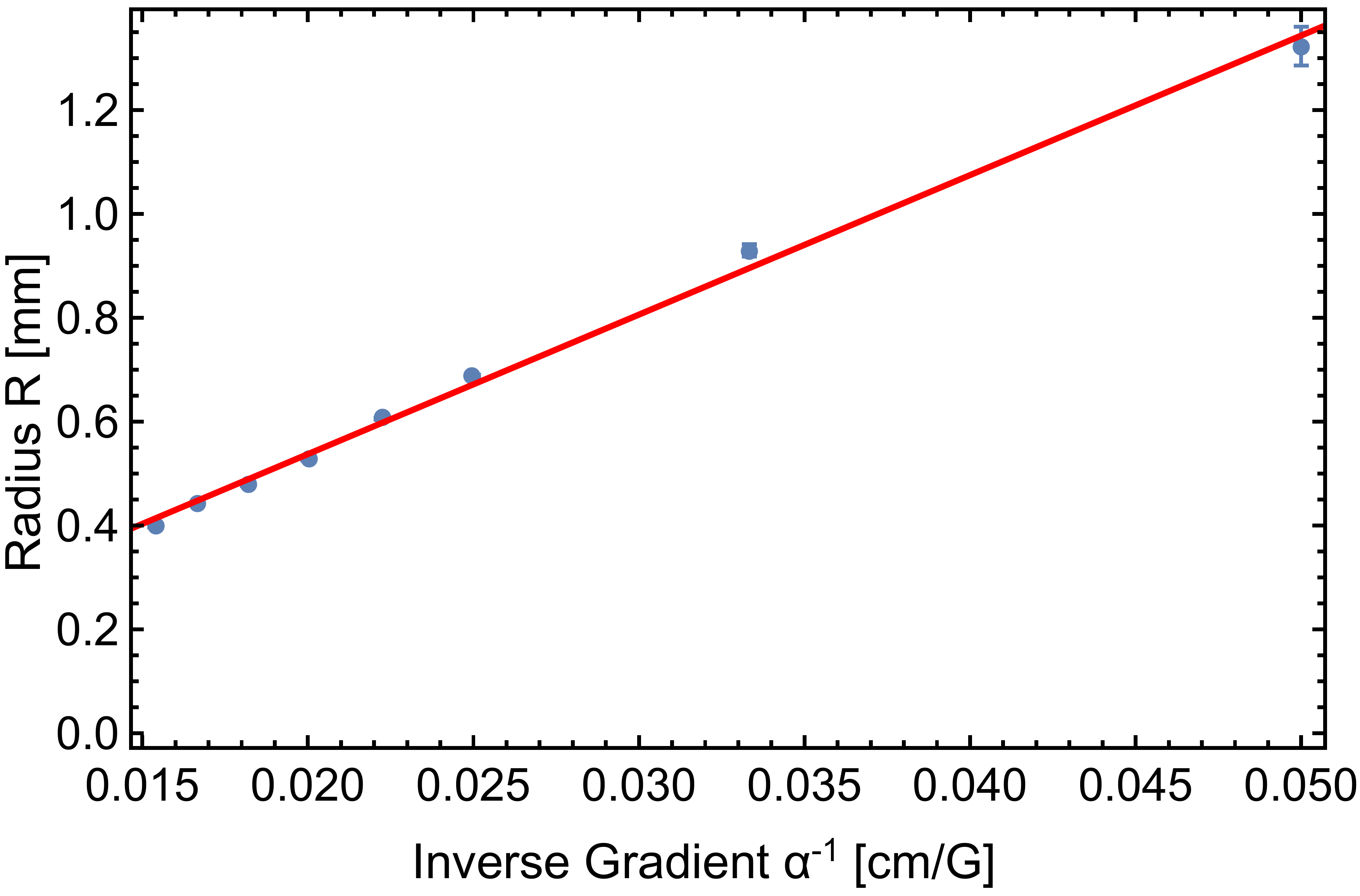}
 \caption{Radius $R$ of the TAAP rings as a function of the inverse magnetic gradient of the quadrupole $(1/ \alpha)$ for a constant rf-frequency of $\omegarf/2\pi=2.5$\,MHz. The polarization of the dressing RF is circular in the x-y plane, the vertical modulation magnitude is 1.9\,G and the Rabi frequency $\Omegarf/2\pi=300$\,kHz. Note that we do not modulate the frequency of the rf, and that therefore the radius of the ring is smaller than one might expect $(R<\hbar \omega_\text{rf0}/g\mu_B \alpha)$. The solid red line serves as a guide to the eye. The error bars correspond to the standard deviation of repeated experiments and is typically $\sim0.3\%$ of the radius.
}
 \label{fig:ringDiameterVsGradient}
\end{center}
\end{figure}

\subsection{The tilted ring trap $(s=0$, $\delta\ne0)$ }\label{sec:tiltedRing}
\begin{figure}[h!]
\begin{center}
 \includegraphics[width=1 \textwidth]{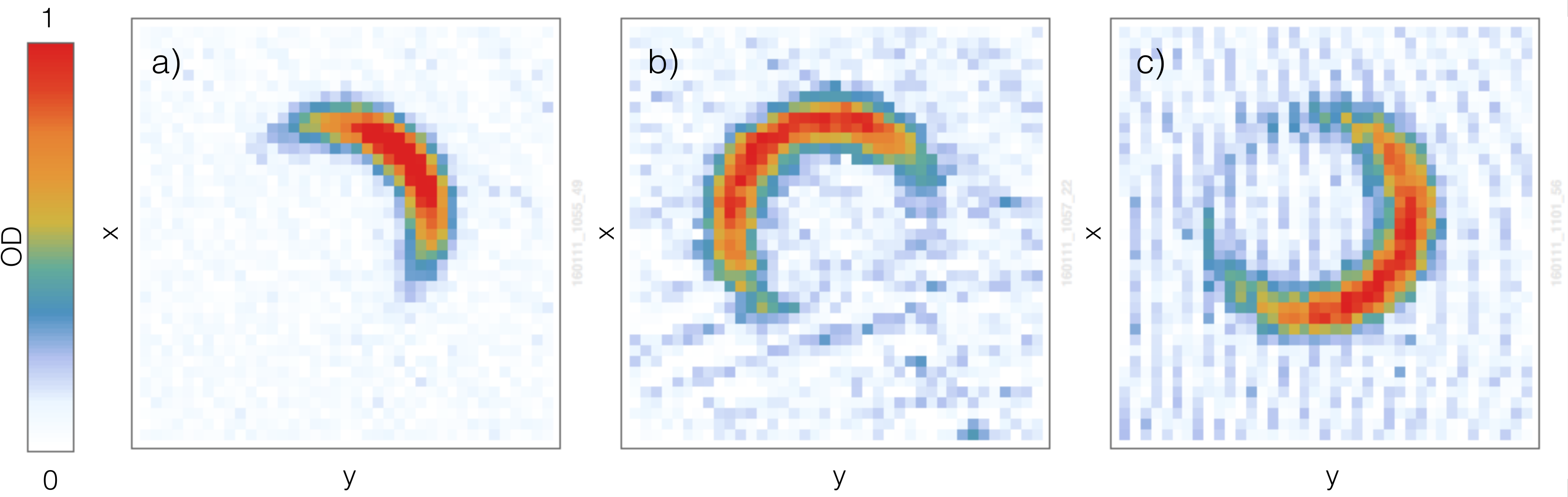}
 \caption{\emph{A tilted ring trap:}
  Experimental realisation of gravity tilted TAAP buckets in $\ket{1,1}$ state ($s=0$) with $\alpha=55$\,G/cm, $\omegarf/2\pi=2.62$\,MHz. The radius is $R=500$\,\textmu m and the atom number is about $3\times10^5$~atoms. In a) $\delta=0.28$\,rad\,$=16^{\circ}$. In b,c) we tilt the ring in the y---z plane, in opposite directions, by the same amount of $16^{\circ}$.}
 \label{fig:ringTilt}
\end{center}
\end{figure}

\noindent The plane of the ring is defined by the direction that the time-averaging field $(B_\text{m})$ is oscillating in. For $\delta=0$ the modulation field is vertical and the ring horizontal. For $\delta\ne0$ the field tilts towards the x-axis and gravity creates a half-moon shaped trap in the direction of the tilt (see Figure~\ref{fig:ringTilt}).
For $\delta \ll 1$, the combination with the gravity field shifts the center of the z-confinement according to $z \rightarrow z'=z-R\delta \cos \phi/2 -g/\omega_\text{z}^2$, where $\phi$ is the azimuthal angle.
By slightly tilting the modulation field ${\bf B}_\text{m}$ from the purely vertical ($0<\delta\ll1$) we can tilt the ring  thus creating a crescent shaped trap with a potential of

\begin{equation}\label{eq:tiltRing} 
V_\text{g}^\text{eff} ({\bf r})=V_\text{ring}^\text{eff} (r,z') - \frac{mg^2}{{2}\omega_\text{z}^2} - \delta\frac{1}{2}m g R \cos(\phi). 
\end{equation} 
Figure~\ref{fig:ringTilt}(a) shows an experimental realisation of a tilted trap of 1\,mm diameter with $3\times 10^5$ Rubidium atoms in the $\ket{1,1}$  state with $\delta=0.28$\,rad and Figs.~\ref{fig:ringTilt}(b),(c) are ring traps with the same number of atoms and trap conditions, tilted in two opposite directions along the y---z plane. 

\subsection{The state dependent trap $(s\ne0$, $\delta=0)$}\label{sec:stateDependent}
\begin{figure}[h!]
\begin{center}
 \includegraphics[width=0.8 \textwidth]{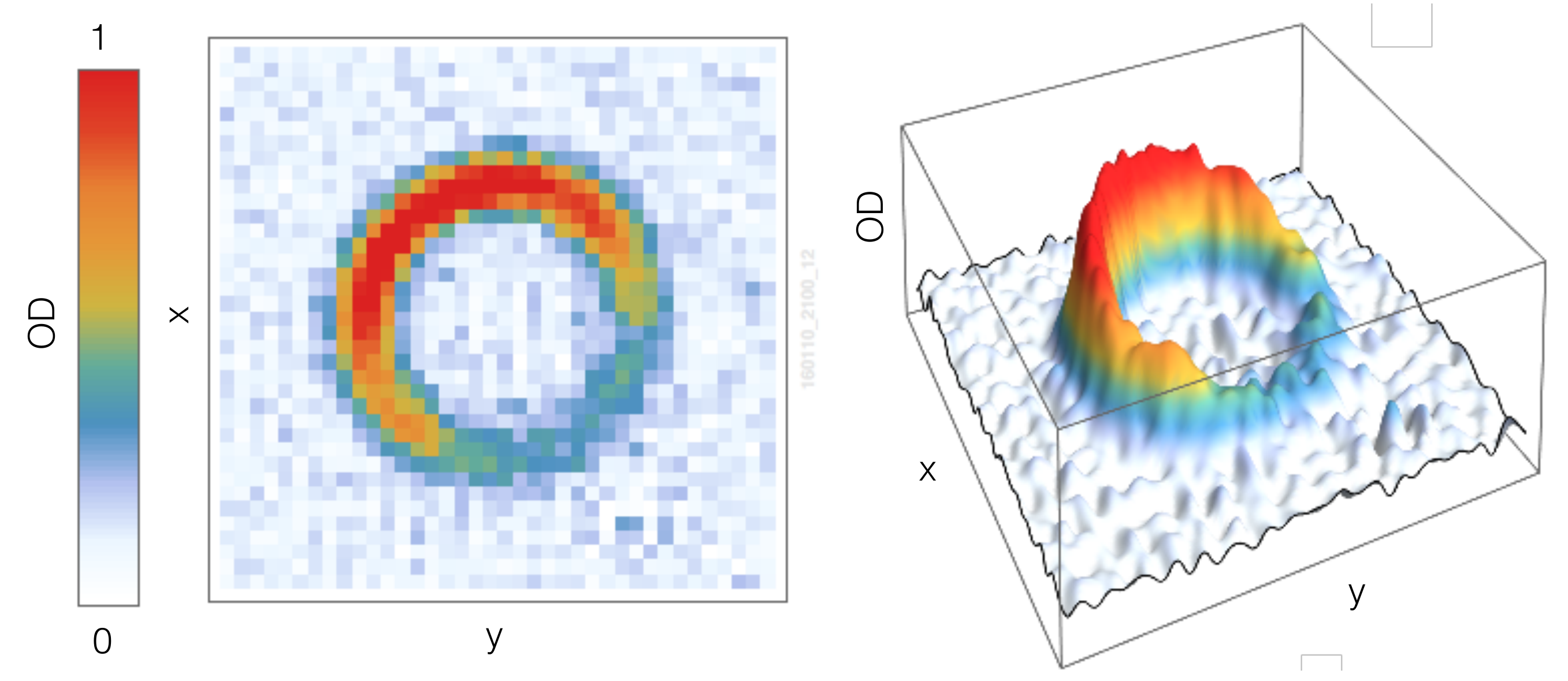}
 \caption{Experimental realisation of a state dependent TAAP bucket $(s\neq0$,\,$ \delta=0)$ for the  $\ket{2,2}$ state at a quadrupole gradient of $\alpha=55$\,G/cm and $\omegarf/2\pi=2.62$\,MHz. $3\times10^5$ atoms were trapped in the bucket and the radius is $R=490$\,\textmu m.}
 \label{fig:ringState}
\end{center}
\end{figure}

\noindent The Rabi coupling for circular rf polarisation depends on the selection rules for the transition and thus on the state of the atoms. The pure ring trap of Section \ref{sec:pureRingTrap} has a strong vertically polarized rf field. 
Adding to it a small, circularly polarized rf will increase or decrease the coupling strength $\Omegarf$ of Eq.~\ref{eq:coupling} depending on which state the atoms are in (e.g.~see Fig.~\ref{fig:ringState} for the particular demonstration of a state dependent TAAP bucket in the $\ket{2,2}$ state). 

For a small positive $s$, the circular rf component lies in the x-z plane and increases the Rabi frequency of the $\ket{2,1}$ state in the positive y-direction while at the same time decreases it for the $\ket{1,1}$ state. 
This creates a trap located at ${\bf r} = (-R,0,0)$ for the $\ket{2,1}$ state and a trap located at ${\bf r} = (+R,0,0)$ for the $\ket{1,1}$ state. 
For a slightly elliptic rf $(0<s\ll 1)$ we obtain to first order:
\begin{equation}\label{eq:StateDependentRing}
V_{\pm}^\text{eff} ({\bf r})=V_\text{ring}^\text{eff} (r,z)
\pm\hbar \Omega_{0c}\frac{2s}{\pi}E(i\beta)\sin \phi\,,
\end{equation}
where $E(x)$ is the complete elliptic function. 

\subsection{Arbitrary traps $(s\ne0$, $\delta\ne0)$}\label{sec:ArbitraryTraps}
\begin{figure}[h!]
\begin{center} 
 \includegraphics[width=0.9 \textwidth]{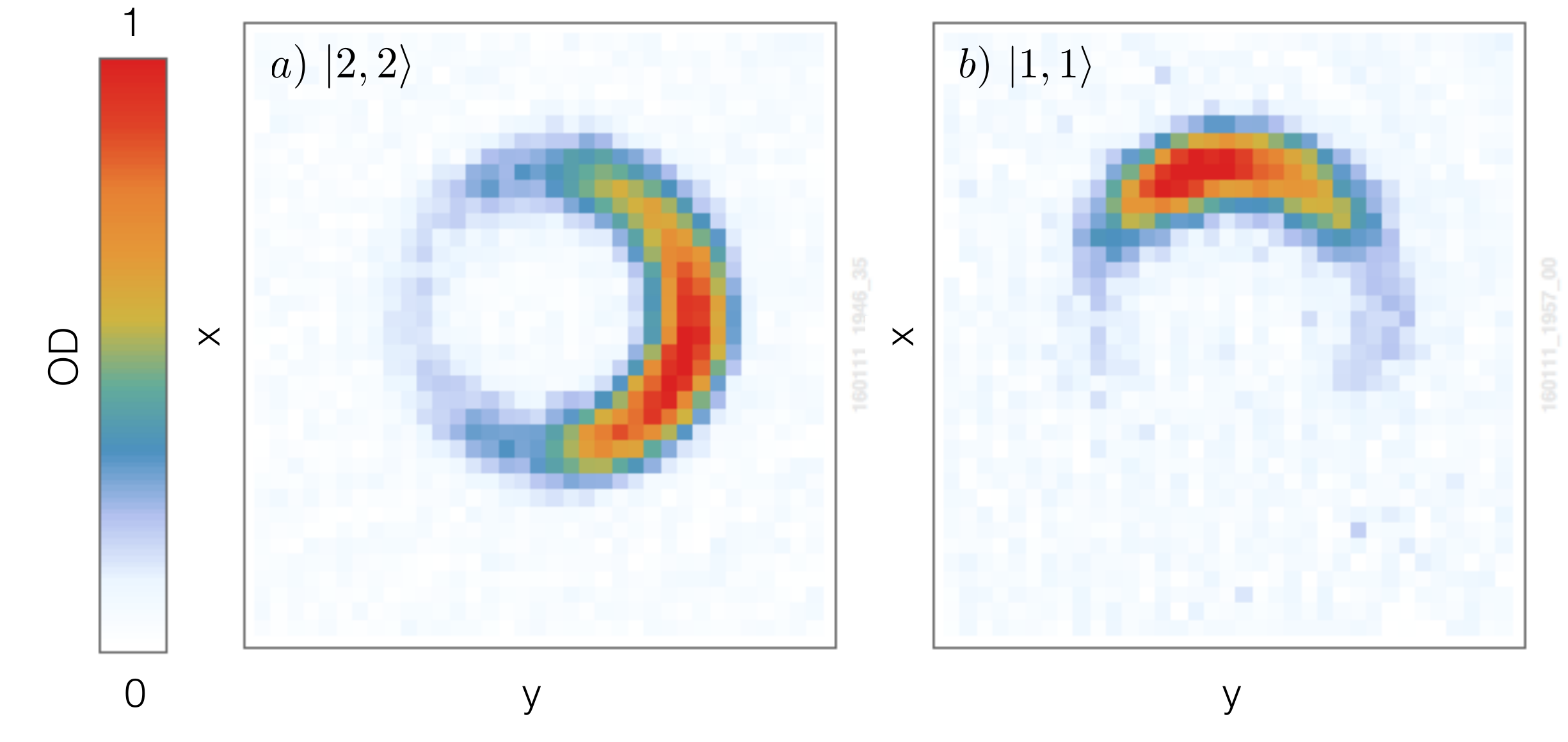}
 \caption{Experimental realisation of  arbitrary traps with $(s\neq0$,\, $\delta \ne0)$ for the two states $\ket{2,2}$ ($5\times10^5$ atoms) and $\ket{1,1}$ ($3\times10^5$ atoms) at $\omegarf/2\pi=2.62$\,MHz. 
 The fitted radius is $440$\,\textmu m and $450$\,\textmu m respectively. The quadrupole gradient is $\alpha=55$\,G/cm. Note that a) and b) are taken with identical experimental conditions and differ only in the state of the atoms. The axis of the circular rf component and the one of the tilted modulation are not orthogonal.}
 
 \label{fig:ArbRing}
\end{center}
\end{figure}

\noindent Combining  the state dependent potentials $(s\ne0)$ with a tilt of the modulation field $(\delta\ne0)$, we  create two independent traps each at an arbitrary position along the ring (e.g.~see Figure~\ref{fig:ArbRing} for a demonstration with the $\ket{1,1}$ and $\ket{2,2}$ states)

The resulting potential is then:

\begin{equation}
\begin{aligned}\label{eq:fullPot}
V^\text{eff}_\pm ({\bf r})&=V_\text{ring}^\text{eff} (r,z') - \frac{mg^2}{{2}\omega_\text{z}^2}- V_0 \cos(\phi \mp \phi_0)\,\,,
\\
V_0&=\sqrt{\left(\frac{2}{\pi}E(i\beta)\hbar \Omega_{0c}s\right)^2+ \left({\frac{1}{2}}m g R\delta\right)^2}\,\,,
\\ \tan(\phi_0)&=\frac{4s\hbar \Omega_{0c}E(i\beta)}{\pi m g R\delta}.
\end{aligned}
\end{equation}
 
By adjusting $s$ and $\delta$, we can adjust the trap position $\phi_0$ to take any value between $0$ and $2\pi$. 

The two traps are anti-symmetric with respect to the x-z plane.
Note also that in principle the tilt of the modulation and the weak circular rf component can point in any direction.

The corresponding azimuthal trap frequency 
\begin{equation}
 \omega_\phi=\frac{1}{R}\sqrt{\frac{V_0}{m}}\,. \label{eq:trapFreuencyFi}
\end{equation}
governs the speed at which one can move the two traps around the ring without exciting the trapped atomic cloud.

Figure \ref{fig:ArbRing} shows a TAAP with two traps in arbitrary positions.
Both images are taken at identical experimental conditions, but with different states: in figure \ref{fig:ArbRing}a the 
atoms are in the $\ket{2,2}$ state and in figure \ref{fig:ArbRing}b  in the $\ket{1,1}$ state. 
The axis of the circular rf component and of the tilt are shifted by $\pi/4$.

\subsection{Landau Zener losses}

From equation (\ref{eq:trappingFrequencies}) it is clear that one has an interest in reducing $\Omega_0$ and thus $B_\text{rf}$ in order to achieve a maximal radial confinement.  
This however is limited by Majorana spin flips, which occur when the rf-coupling between the energy levels is too weak for the atoms to follow the adiabatic states. 
According to the Landau-Zener criterion the probability of such transition is given by $P=\exp(-2\pi \Gamma_\pm)$. Since the time variation of the potential experienced by the atoms depends mainly on the modulation, rather than the atom's motion, we obtain according to~\cite{Lesanovsky2007PRL}:

\begin{equation}
\Gamma_\pm({\bf r},t)=\hbar\frac{\Omega_{\pm}^2({\bf r},t)}{\partial_\text{t}\left(\Omega_\text{L}({\bf r},t)-\omegarf(t)\right)}\gg 1
\end{equation}
Assuming a small tilt $\delta \ll 1$  and a low amplitude circular rf $s \leq 1/\sqrt{1+\beta^2}$ then spin flip losses are suppressed when:
\begin{equation}
\omega_\text{m} ~\ll~\frac{\Omega_{0c}^2 R}{\,\omega \, a_\text{ho}}\,\,,
\end{equation}
where $a_\text{ho}=(\hbar/m\omega_\phi)^{1/2}$ is the azimuthal harmonic oscillator length.

\section{TAAP Interferometers} \label{sec:TAAP_Interferometers}

Here we discuss the different types of matter-wave  interferometers made possible by the TAAP potentials. 
We  start with  interferometers using the matter-wave guides of section \ref{sec:pureRingTrap} and introduce two different types of beam splitters: one based on Bragg beams (section \ref{sec:BraggRingInterferometer}) and  one based on rf/microwave two photon transitions and state selective manipulation of the atoms (section \ref{sec:StateDependentInterferometer}).
We  then show that the latter scheme can also be used in the fully trapped case (section \ref{sec:movingBucket}).
In order to avoid any contribution other than the Sagnac phase $(\phi^0_{\text S})$, we ensure  that the state-dependent trajectories of the atoms are perfectly symmetric for $\Omega=0$. 
We achieve this by constructing the interferometric sequence such that every point in the trajectory is traversed by the two states in exactly the same internal state. 
The only difference between the trajectories of the two clouds is then the direction in which the atoms travel.

\subsection{The Sagnac phase} \label{sec:InteferomtryTheoryMain}

We will shortly review the Sagnac phase as it applies to our guided interferometers.
The Sagnac phase occurs in atom trajectories that enclose an area. 
It is often assumed that the two wave packets overlap perfectly at  the second beam-splitter. 
Insufficient overlap is usually taken into account only as a reduction in contrast.
For coherent matter-waves such as those originating from a condensate one can achieve a large contrast even when the  overlap is limited \cite{Andrews1997S, Jo2007PRL2}.
We will show, that in this case there is nevertheless a correction to the Sagnac phases. 

\begin{figure}[h]
\begin{center}
\includegraphics[width=5cm]{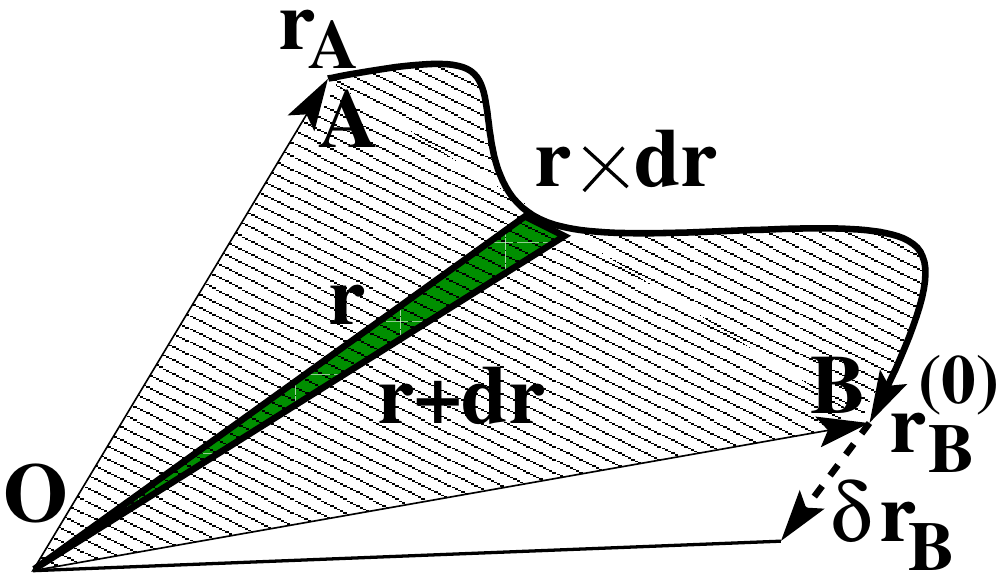}
\caption{Representation of the area covered by the atoms together with the correction
to non adiabaticity.}
\label{fig3b}
\end{center}
\end{figure}

This can be formulated under the following rather general theorem:
We assume non interacting atoms evolving from point $A$ to $B$ with phase accumulation given by
$\Phi_\pm(t)$.

\begin{figure}
\begin{center}
\includegraphics[width=4cm]{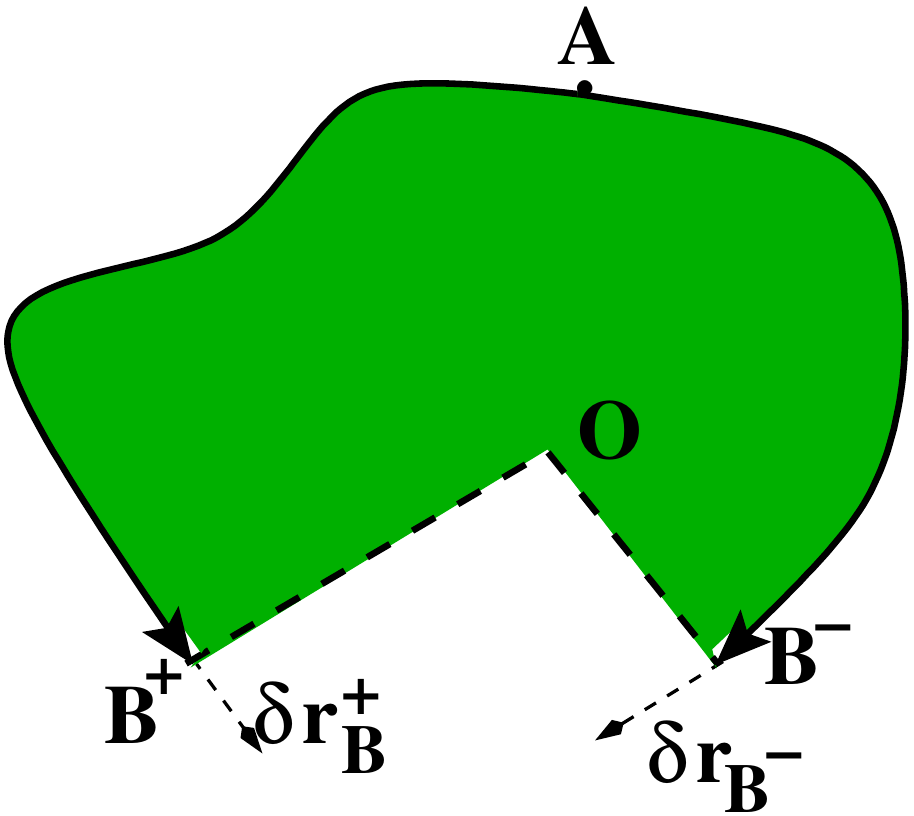}
\caption{The `area` encircled by the atoms of the two states if the atom trajectories $\bf B^-AB^+$ are not entirely closed.  $O$ represents the origin of the coordinate system and is entirely arbitrary.}
\label{fig:UnclosedArea}
\end{center}
\end{figure}

The starting and end points of the classical trajectory in the absence of rotation are $\bf r_A$ and $\bf r_B^{(0)}$. 
The presence of the rotation changes the  classical Lagrangian with a linear correction  term $\delta L(t)$ that induces the change in the classical action governing the phase accumulation. 
If $\delta \bf r_B$ is the linear correction to the external rotation ${\bf \Omega}$ (see Fig.~\ref{fig3b}), then the linear contribution due to this external rotation is given by: 

\begin{equation}
\begin{aligned}\label{eq:GeneralUncloseArea}
\Phi_{BA}
&= \int_{t_A}^{t_B}\! dt\, \delta L(t)/\hbar
\\
&=\frac{m}{\hbar}\left[ 2{\bf \Omega}.\int_A^B {\bf r}\times d{\bf r}
+ { \bf \dot r}_B^{(0)}.\delta {\bf r}_B\right]\,.
\end{aligned}
\end{equation}
A proof of this relying only on the classical dynamics of the atom motion can be found in Appendix \ref{sec:AppndixBNonAdiabaticInterferometry}. 
The first term of equation\,(\ref{eq:GeneralUncloseArea}) gives the area enclosed by the atoms while the second gives the correction due to non-adiabaticity.
Note that the second term can always be made to vanish by an appropriate choice of origin of the reference frame such that the external rotation motion cancels the final velocity (by adding a term ${\bf \Omega}\times{\bf a}$, where ${\bf a}$ is the vector of the new origin).
The two clouds accumulate two different corrective phases $\delta \Phi_{B^\pm A}$. If we assume that without the external rotation they end up in the same position, the net result is (see also Fig.\,\ref{fig:UnclosedArea}):

\begin{equation}\label{eq:AreaTheoremClosed}
\begin{aligned}
\delta \Phi &= \Phi_{B^+A}-\Phi_{B^-A}
\\
&=\frac{m}{\hbar}\left(2 {\bf \Omega}.\int_{\cal A} {\bf dS}
+ {\bf \dot r}_{B^+}^{(0)}.\delta {\bf r}_{B^+}-
{\bf \dot r}_{B^-}^{(0)}.\delta {\bf r}_{B^-}\right). 
\\
\end{aligned}
\end{equation}
where $\cal A$ is the surface enclosed by the path $B^-AB^+OB^-$ and  ${\bf dS}$ is the differential surface vector. 
Please note that the choice of origin $O$  does not change the value of the integral, since the change in area is exactly compensated by the change in ${\bf \dot r}_{B^+}^{(0)}.\delta {\bf r}_{B^+}-{\bf \dot r}_{B^-}^{(0)}.\delta {\bf r}_{B^-}$.
We retrieve the standard Sagnac formula $\Phi_\text{S}^0= \frac{4\pi\,\Omega {\cal A}}{h/m}$, if at the end of the interferometer sequence the atoms are at rest and the area is closed, i.e.~the atomic clouds overlap fully.

In most Sagnac interferometers the atoms  travel along the identical path in opposite directions.  
In this case the area of the interferometer is the sum of the areas surrounded by each arm and thus twice the area of the single path. 
Taking the particular case of a ring-shaped interferometer of radius $R$ and radial trapping frequency $\omega_\rho$ and  assuming that the atoms are strongly confined to the ring $(\omega_\rho\gg\Omega)$ the  area is  ${\cal A}=2 \pi R^2$. Since ${\bf r}_B^2=R^2$ we find ${\bf r}_{B\pm}^{(0)}.\delta {\bf r}_{B\pm}=0$ and due to the tight confinement  $\delta {\bf r}_{B\pm}=0$.

Therefore, we obtain a much simpler expression 
\begin{equation}\label{eq:IncompleteSagnacPhase}
\begin{aligned}
    	\Phi_\text{S}(T)= \frac{ \delta \phi(T)}{h/m}\Omega {\cal A}= \frac{\delta \phi(T)}{4\pi}\,\Phi_\text{S}^0\,\,, 
\end{aligned}
\end{equation}
where $\delta \phi(T)=\phi_+(T)-\phi_-(T)$ is the difference in angle between the cloud centers accumulated during their trajectory along the ring interferometer, with $\delta \phi (T)=4\pi$ signifying therefore the closed area. 
The Sagnac phase is therefore only affected by the final  position of the mass center of the atoms. 

For a finite radial confinement we have to take into account that the atoms will experience a centrifugal force during their trajectory, which will be compensated by the radial trapping potential. 
The area of the interferometer (and thus its sensitivity) will therefore increase slightly. 
An upper bound for this is 

\begin{equation}\label{eq:tightRadialError}
	\frac{\delta A}{A}\leq\text{Max}\left[2\left(\frac{ \dot{\phi_0}}{\omega_\rho}\right)^2\right].
\end{equation}
Taking a transit time of 1\,s  and  a radial trapping frequency of 400\,Hz, we find a negligible correction of $10^{-5}$. Note that this correction affects only the sensitivity and not the zero-fringe.

Finally, care needs to be taken to minimise the spread of the wavepackets during their trajectory as well as the influence of the interatomic interactions.  For a thermal cloud this can always be done simply by reducing the number of atoms. For a Bose condensed cloud, the phase diffusion time $T_\phi$ can be estimated assuming a Poissonian atom number fluctuations and the chemical potential $\mu$ taken for a Thomas-Fermi profile: 
\begin{equation}
{T_\phi } = \frac{1}{\hbar }\sqrt N {\left. {\frac{{\partial \mu }}{{\partial N}}} \right|_{N/2}} 
= {{\rm{N}}^{ - 10}}{\left( {\frac{{72 a^2\omega_\phi\omega_\rho^2 }}{{125a_{{\rm{Ho}}}^2}}} \right)^{1/5}}, 
\end{equation}
 where $a_\text{Ho}$ is the harmonic oscillator length, $a$ the s-wave scattering length. 
 Surprisingly enough this implies that---as oopposed to the thermal case---the phase coherence time of a BEC interferometer increases for increasing atom number even in the case of absence of squeezing, much in contrast to the thermal case.

\subsection{Guided Sagnac Interferometry}
The basic idea of a Sagnac interferometer is to let atoms travel in opposite directions around an enclosed area, completing a full loop. 
Due to the resulting common path, the interferometer becomes independent of static acceleration (gravity) but will be sensitive to rotation.

The main challenge for \textit{guided} Sagnac interferometers is the smoothness of the waveguides. In the case of magnetic guiding potentials the shape of the waveguide is defined by the shape of the field generating wires. 
Therefore, imperfections in the wires translate directly into imperfections of the waveguides. 
This is the main reason why  interferometry in waveguides  has only been demonstrated over microscopic distances so far \cite{Wang2005PRL,Schumm2005NP,Ockeloen2013PRL}. 
The TAAPs of this paper are generated from coils that are large and far away. The resulting waveguides are therefore  inherently smooth (see section \ref{sec:TAAPs} and \cite{Lesanovsky2007PRL}). 

A second challenge is the beam splitter. As will be discussed in section \ref{sec:BraggRingInterferometer}, 
the standard beam splitters of atom interferometry (Raman or Bragg), can also be used in TAAP interferometers. 
A completely different beam splitter will be discussed in section~\ref{sec:StateDependentInterferometer}, where we propose to use state-dependent manipulation of a superposition of atoms in the rf-dressed $\ket{F,m_\text{F}}=\ket{1,1}$ and $\ket{2,1}$ states. 

With an appropriate choice of parameters the linear Zeeman energy can be nearly identical for both states~\footnote{By carefully choosing the trapping parameters, the quadratic field dependence of the energy difference between the two states can also be eliminated.
This is the case for a static magnetic field at 3.23~Gauss \cite{Harber2002PRA}. Similar results have been predicted for rf-dressed potentials \cite{Kazakov2015PRA}. The ideal dressing and modulation fields of the TAAPs will be investigated separately.}. 
However, they interact differently with different polarisations of the rf dressing fields.
A $\sigma_+$ polarized rf field will couple the $\ket{1,-1}$ to the $\ket{1,+1}$ state, but not the $\ket{2,+1}$ to the $\ket{2,-1}$ state. The opposite is true for a $\sigma_-$ polarized rf. 
We can therefore use the \emph{degree of circular polarisation} of the rf dressing field in order to manipulate the atoms in a state-dependent fashion.

Combining an elliptically polarized rf-field with a tilt of the ring-shaped waveguide, we can move the two states independently to anywhere on the ring (e.g.~Fig.~\ref{fig:ArbRing}) including moving the two clouds  around the ring in opposite directions.
For such a case the standard Sagnac phase is defined as 
\begin{equation}\label{eq:SagnacOriginal}
	\Phi_\text{S}^0= \frac{4\pi\,\Omega {\cal A}}{h/m},
\end{equation}
where $\Omega$ is the angular velocity of the rotation to be determined, $m$ the mass of the atom and $\cal A$ is the area enclosed by the arms of the interferometer \cite{Barrett2014CRP,Antoine2003PLA}. 
For the case of the ring-shaped wave guide discussed in this paper, where the two wavepackets travel once around the ring in opposite directions, the area is $A=2\pi \rho$, where $\rho$ is the radius of the ring. 

The Sagnac formula still holds both for wave-guided interferometers and 3D-trapped atomic clouds. 
We extend the concept of the area to cases where the paths are not fully closed and consider corrections to the area taking into account the centrifugal forces. 
As shown previously, the phase accumulation corresponds to the classical action (in units of $\hbar$) taken over the trajectory \cite{Antoine2003PLA}.

\subsection{A waveguided interferometer with a Bragg beam splitter}\label{sec:BraggRingInterferometer}

Section \ref{sec:pureRingTrap} demonstrates the ring-shaped TAAP trapping potentials $(s=0$, $\delta=0)$, that will serve as waveguides for the interferometer. In order to place an atom cloud into this waveguide we load a cold thermal cloud or a BEC into the harmonic trap created by tilting the modulation fields 
$(s=0$, $\delta\ne 0)$ as discussed in section \ref{sec:tiltedRing}. We then switch suddenly to the flat ring  $(s=0$, $\delta=0)$ and apply a near resonant Bragg pulse, which splits the atomic cloud into two momentum states $\pm 2 \hbar k$ in the laboratory frame, with $k=2\pi/\lambda$ being the wave vector of the light \cite{Giltner1995PRL}.
The atoms then move along the ring with a constant (angular) velocity ${\dot \phi}_{0\pm}(t)=\pm 2\pi /T$, where $T=2\pi R /(2 v_\text{rec})$ is the half round trip time and $v_\text{rec}$ is the recoil velocity of the Bragg beams.
After a full trip around the ring we apply a second Bragg pulse, which then produces three atomic clouds: one at rest $p=0$ and two with a momentum of $p=\pm 2 \hbar k$. 
The interferometric signal is then calculated by subtracting the sum of the populations of the $p=\pm 2 \hbar k$ wave packets from the population of the  $p=0$ wave packet. 
We assume tight radial and axial confinement in the ring, so that the atoms move according to  ${\bf r}_{0\pm}(t)= \big(R\cos[(\pm 2\pi/T+\Omega) t], R\sin[(\pm 2\pi/T+\Omega) t],0\big)$. 
Since the atoms remain at the trap minimum only the kinetic energy contributes  to the phase difference. We obtain a phase shift for a round trip of the two clouds:

\begin{equation}
\Phi_\pm(T)=\frac{mR^2}{2}\int_0^T \frac{dt}{\hbar}\left(\Omega\pm \frac{2\pi}{T}\right)^2.
\end{equation}
By calculating the phase difference between the two trajectories we recover the standard Sagnac phase of equation (\ref{eq:SagnacOriginal}):
$\Phi_\text{S}(T)=\Phi_+(T)-\Phi_-(T)=\Phi_\text{S}^0$. In the more general case, however, the components have different  angular velocities $\phi_\pm(t)={\dot \phi}_{\pm}t$, and thus ${\bf r}_{0\pm}(t)= \big(R\cos[({\dot \phi}_{\pm}+\Omega) t], R\sin[({\dot \phi}_{\pm}+\Omega) t],0\big)$. An example is the use of Raman beamsplitters, where the two beams may have different frequencies.
This results in the modified Sagnac signal: 
\begin{equation}\label{eq:SagnacSquared}
\Phi_\text{S}=
\frac{2 \pi R^2T}{h/m}\left[({\dot \phi}_+-{\dot \phi}_-)\Omega
+({\dot \phi}_+^2-{\dot \phi}_-^2)/2\right]
\end{equation}

Raman pulses in free space offer very precise control over the difference in the velocity of the split atomic clouds. 
For the case of a TAAP waveguide, however, the alignment of the potential with respect to the optical Raman beams has to be taken into account.
In practice this means that the square terms in the Sagnac equation (\ref{eq:SagnacSquared}) severely limit the precision in the determination of $\Omega$. 
It is therefore desirable to choose the Bragg scheme, where the two velocities are identical and the non-linear terms vanishes. 

\subsection{Interferometry based on state dependent manipulation of atoms in a ring}\label{sec:StateDependentInterferometer}

Even though the simplicity of the Bragg scheme is conceptually very appealing, it has the major drawback that the reference for the Sagnac phase is the standing wave of the Bragg beams and thus the retro-reflecting mirror which generates it. 
Whereas in free-space interferometers this is a very suitable reference, in wave-guided interferometers this poses the problem that this can cause a spurious phase shift when  the position of the wave guide moves relative to the one of the retro-reflecting mirror. 
We therefore have to look for interferometers, where the phase reference is defined by the trapping fields themselves. 
This is the case for clock type interferometers \cite{Stevenson2015PRL} using as a  beam splitter a two photon rf/microwave pulse. 
The position of the beam splitter is then simply the position of the atoms in the trap.

We  start with atoms in the $\ket{2,1}$ state in the tilted ring described in section \ref{sec:tiltedRing} at the initial angular position $\phi_\pm=\phi_0(t=0)=0$. 
The atoms are at rest in the rotating frame, therefore the initial velocity in the inertial frame is $\Omega R$. 
We then  apply a two-photon $\pi/2$ pulse, which puts the atoms into an equal superposition of the two dressed hyperfine states $\ket{2,1}$ and $\ket{1,1}$. We then move the atoms half way around the ring\,\footnote{Note that the experimental conditions are such that these two components experience the same trapping potential.}.
After half of a round trip we apply a $\pi$ pulse, which turns the $\ket{2,1}$  state into the $\ket{1,1}$ state. 
We then complete the round trip for each of the two states and apply a second $\pi/2$ pulse. Due to the $\pi$ pulse the experimental sequence of the second half can be exactly the time reversal of the first half, thus ensuring perfect symmetry of the whole interferometric sequence. There are two principle options to move the atoms around the ring interferometer: a) the atoms are subjected to an initial state-dependent acceleration and then propagate freely in a TAAP matter-wave guide the static or dynamic \emph{accelerator rings} described below or b) the atoms are fully confined in three dimensions using the \emph{moving buckets} of section~\ref{sec:movingBucket}.

\subsubsection{The Accelerator Ring:}\label{sec:StaticAccelerator}

After preparing the atoms in a superposition of $\ket{2,1}$ and $\ket{1,1}$, we switch to a  state-dependent trap $(s\ne0$ and $\delta=0$, as discussed in section \ref{sec:stateDependent}) with the two trap centers at $\phi_{0\pm}=\pm\pi/2$. 
The $\ket{2,1}$ and $\ket{1,1}$  components of the atom clouds then start to accelerate symmetrically in opposite direction until they reach  $\phi_{0\pm}=\pm\pi/2$, at which point they start to slow down.
Due to the symmetry of the potential the clouds come to a standstill after they have concluded a half turn $(\phi_{0\pm}=\pi)$. 
We then apply a $\pi$-pulse, which turns $\ket{2,1}$ into $\ket{1,1}$ and vice versa. 
The atom clouds, now with their identity exchanged, accelerate again and continue to travel in the same direction as before, until they reach $\phi_{0\pm}=\mp\pi/2$ and finally come to a standstill at $\phi_{0\pm}=0$.
A further $\pi/2$ pulse then closes the interferometer, with the readout being the difference in populations of the $\ket{2,1}$ and $\ket{1,1}$ states.
Conveniently, these two components separate again and are easily imaged when they reach their maximum separation at $\phi_{0\pm}=\pm\pi/2$. The angle $\phi_\pm(t)=\pm \phi(t)$ follows the nonlinear classical equation (see Appendix B):
\begin{equation}\label{eq:nolin}
mR^2{\ddot \phi}(t)=V_0(t)\sin(\phi_0(t)- \phi (t))
\end{equation}
and formula (\ref{eq:IncompleteSagnacPhase}) still holds.

Since the TAAP waveguides are extremely smooth \cite{Lesanovsky2007PRL}, no transverse modes are excited during the transit and equation (\ref{eq:nolin}) yields
\begin{equation}\label{stat} T=2\sqrt{\frac{mR^2}{2V_0}} \int_0^{\pi} \frac{d \phi}{\sqrt{\sin \phi}}\simeq 7.4/\omega_\phi. 
\end{equation} 
With the atoms overlapping and being at rest during the second $\pi/2$ pulse, the interferometer is well described by the Sagnac phase $\Phi_{\text S}$.

The fact that the atoms come to a standstill at $\phi_{0\pm}=\pi$ causes an unnecessarily slow transfer of the atoms. 
The transit time can be  reduced by dynamically adjusting the arbitrary ring potentials of section \ref{sec:ArbitraryTraps} such that  the acceleration in the first half of the cycle and the deceleration in the second half both remain maximum. Care should be taken for the relative velocity of the two clouds to remain small or densities sufficiently low so as to avoid s-wave scattering \cite{Buggle2004PRL}.

\subsubsection{The moving buckets:}\label{sec:movingBucket}

In most inertial atom-interferometric measurements the atoms are in free fall for most of the interferometer sequence.
Even in the case of the waveguided interferometer described above the atoms are free to travel around the ring.
However, as is evident e.g.~from equation~(\ref{eq:AreaTheoremClosed}) and described in \cite{Stevenson2015PRL,Fernholz2007PRA}, this is not a strict requirement: atoms in a matter-wave guide or even in a 3-D trap still reproduce the Sagnac phase.
Even macroscopic atom clocks flown once around the globe display the Sagnac phase \cite{Hafele1972S}.
In the following sections, we describe Sagnac interferometers where the atoms are for most of their trajectories fully 3D-trapped.
One important advantage of this is that this prevents the atomic cloud from expanding excessively during its trajectory.
 
\begin{figure}[h]
\begin{center} 
 \includegraphics[width=9cm]{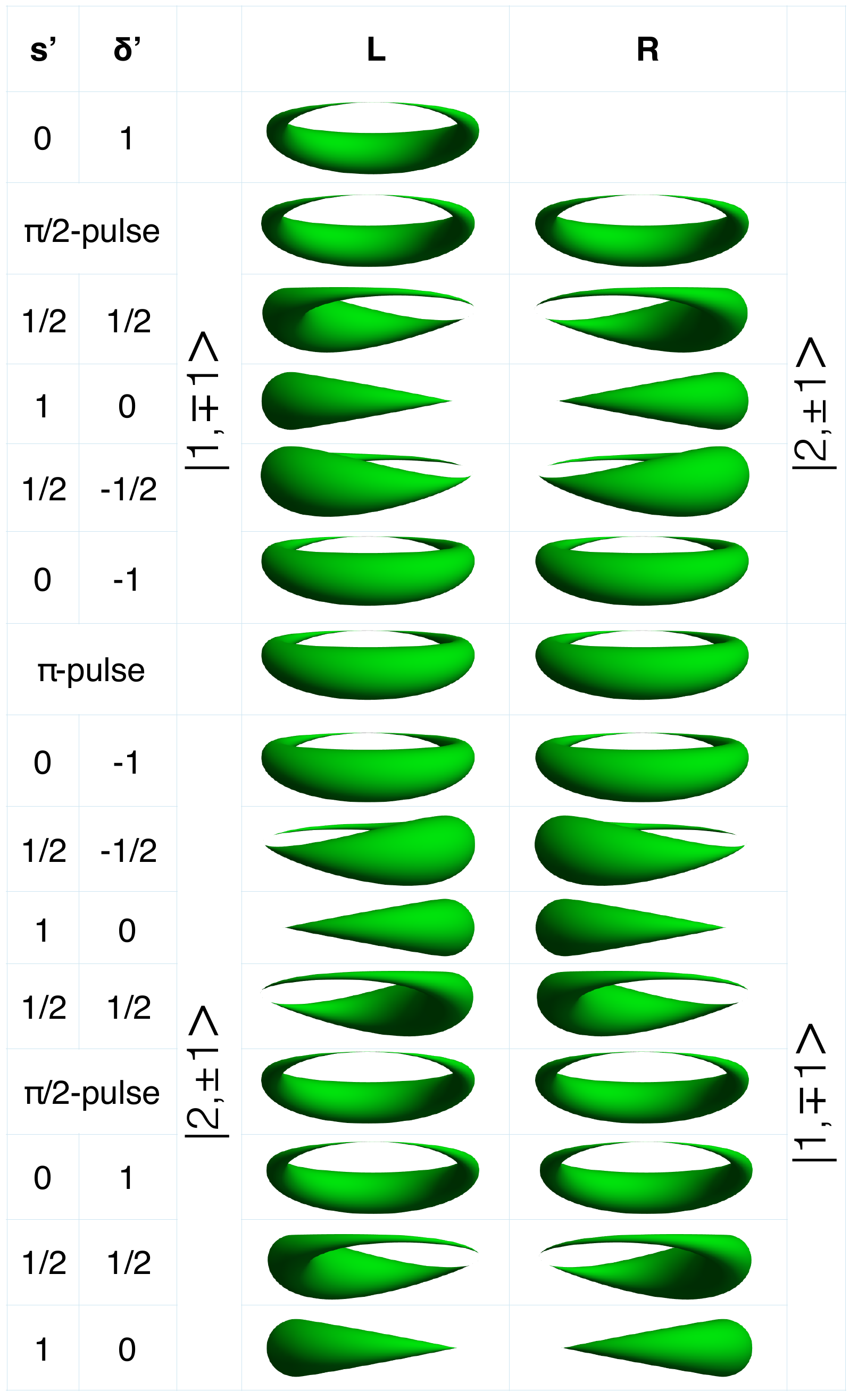}
 \caption{A sketch of the isopotential surfaces for the `bucket' traps showing the different stages of transport during the interferometer sequence. $\bf s'$ and $\bf \delta'$ stand for the degree of tilt and circular polarisation respectively.}
 \label{fig:BucketPotentials}
\end{center}
\end{figure}

We start with atoms in the $\ket{2,1}$ state at the initial angular position $\phi_\pm=\phi_0(t=0)=0$ and initial radius $R$. The atoms are at rest in the rotating frame, therefore the initial velocity in the inertial frame is $\Omega R$. 
We  apply a two-photon $\pi/2$ pulse, which puts the atoms into an equal superposition of the two dressed hyperfine states $\ket{2,1}$ and $\ket{1,1}$. 
The two states are then moved in opposite directions for a half of a turn around the ring ($\phi_\pm(t=T)=\pm \pi$). 
When the atoms are exactly at the opposite side of the ring, i.e.~after a transit time $t=T/2$ we apply a $\pi$-pulse, which swaps the two hyperfine states.
When the full turn is complete ($t=T$) the atomic clouds are recombined by a second $\pi/2$ pulse at $\phi_\pm(t=T)=\pm 2\pi =0$. 
Note that each of the two wave packets travel through the exactly same regions of space in the same hyperfine state. 
In a static inertial frame, the trajectory of the atoms is therefore symmetric with respect to time reversal and with respect to the plane spanned by $\phi=0$ and the z-axis. 
The interferometer is therefore not sensitive to any static (inhomogeneous) external field or acceleration (gravity). We realize three dimensional trapping using a combination of the state-dependent and gravitational potentials as described in equation\,(\ref{eq:fullPot}). The position of the trap minimum angle $\phi_0(t)$ expressed in equation \,(\ref{eq:fullPot})  is tuned slowly  by a variation of the ellipticity ($s$ parameter) and the rotation ($\delta$ parameter) of the modulation field so that the condensates follow the trap center.

Let us first consider an adiabatic transfer of the atoms around the ring. In order for the atoms to remain at the ground state of the trap, the trap frequencies have to be much higher than the typical time variation associated with $\phi_0(t)$, i.e.~$ \ddot{\phi_0}(t) << \omega_\phi^2$. 
The position of the atoms is then ${\bf r}_\pm(t)={\bf r}_{0\pm}(t)$ and the equations\,(\ref{eq:pos}-\ref{eq:lag})  yield ${\bf \dot r}_\pm(t)={\bf \dot r}_{0\pm}(t)=\hbar{\bf k}_{\pm}(t)/m $. 
Assuming a slow rotation $\Omega$, tight radial confinement, and that the final velocity of the two atom clouds is zero $({\dot \phi_0}(T)=0)$ we get the usual Sagnac phase shift $\Phi_\text{S}(T) =\Phi_\text{S}$.

In a real experiment the adiabatic transfer would require an extremely slow acceleration of the trap ($R \ddot{\phi_0}(t) << \omega_\phi^2$) and a very long transit time $(T\ll2\pi/\omega_\phi)$.
A faster motion would cause the center of mass of the atoms to oscillate with respect to the center of the trap.
However, for a harmonic trap this oscillation can be optimally controlled \cite{Guery-Odelin2014PRA}

Let us now look at the case when the trap moves too fast for the adibaticity to be preserved, but slow enough for the atoms to remain within the harmonic limit of the trap. For a trap moving at a constant speed $\phi_{0\pm}(t)=\pm2\pi\, t/T$ and assuming the atoms start in the laboratory frame at position $\phi(0)=0$ and are initially at rest ($\dot\phi(0)=0$) we can write down the classical motion of the atoms as
\begin{equation}
\begin{aligned}\label{eq:phit}
    \phi_\pm(t)&=\pm2\pi\left[\frac{t}{T} -\frac{\sin(\omega_\phi t)}{\omega_\phi T}\right]\\
    \dot\phi_\pm(t)&=\pm\frac{2\pi}{T}\left[1-\cos(\omega_\phi t)\right] 
\end{aligned}
\end{equation}
If we assume tight radial confinement then equation (\ref{eq:IncompleteSagnacPhase})  yields the Sagnac phase 
\begin{equation}
\label{eq:yanharm}
\Phi_\text{S}(T)
=\left[1-\frac{\sin(\omega_\phi T)}{\omega_\phi T}\right]\Phi^0_\text{S}
\end{equation}
Therefore, we recover the standard Sagnac phase $(\Phi^0_\text{S})$ if the transit time $T$ is a multiple of the trap oscillation time $(2\pi/\omega_\phi)$.

We can also avoid the oscillation by starting with static traps at the position $\phi_{0\pm}(0)=\pm\phi_\text{a}$ rather than   at $\phi_0(0)=0$. 
A quarter of oscillation period later $t=\pi/2\omega_\phi$ the atoms will then be exactly at the trap bottom and move at an angular velocity $\dot\phi_\pm=\pm\omega_\phi \phi_\text{a}$. If at that moment we start to move the trap at the same angular velocity, then the atoms are transported in the ground state of the trap. Stopping the trap at $\phi_{0\pm}=\mp\phi_\text{a}$ and waiting a further  quarter of an oscillation period  perfectly overlaps the two atom clouds at zero relative velocity. We therefore measure the standard Sagnac phase.  The transit time for this trajectory is then $T=\frac{2\pi}{\omega_\phi}(1/2-1/\pi+1/\phi_\text{a})$. This trajectory guarantees that for most of the transit the atoms are at rest in the moving frame and is much faster than the adiabatic requirement $(T\ll2\pi/\omega_\phi)$.

\subsection{Sensitivity}\label{sec:Sensitivity and discussion}

We have demonstrated rings with radius' ranging from $R=200$\,\textmu m to 1.5\,mm. 
Using equation (\ref{eq:SagnacOriginal}) with ${\cal A}=2\pi R^2$ we calculate the transfer function or scale factor of  a ring-interferometer with a radius $R=1.5$\,mm to be $\Phi_{\text S}/\Omega=4\times 10^{4}$\,rad/(rad/s).
The earth's rotation of $\sim 72$\,\textmu rad/s therefore results in a  phase shift of $\Phi_{\text S}\simeq 3$\,rad.

Numerical simulations for a condensate in our experimental conditions indicate that the Sagnac phase can be preserved for atom numbers as large as a few thousand \cite{Trombettoni2015}, resulting in a shot noise limited phase resolution of tens of milli radians. 
Given our repetition rate of a few tens of seconds, we expect a sensitivity of $10^{-5}$\,rad/s/Hz$^{1/2}$. According to \cite{Stevenson2015PRL} also thermal atoms can be used. In that case, using  $10^6$ thermal atoms at a repetition rate of 1\,Hz and pushing the ring diameter to one centimeter a sensitivity of $10^{-9}$\,rad/s/Hz$^{1/2}$ will be reached.
 
\section{Conclusions}\label{sec:Conclusions}

We demonstrated and analysed ring-shaped potentials for ultra-cold atoms and propose different schemes for its use in Sagnac interferometry.
We demonstrated that using an elliptical polarization of the rf-field in combination with gravity we can arbitrarily and dynamically manipulate two different spin states at the same time.

We proposed a number of different Sagnac interference experiments based both on matter-wave guiding and fully trapped manipulation of the atoms.  
We analysed the contributions to the Sagnac phase due to dynamic effects, and found that the standard formula also holds for non adiabatic situations provided the atomic clouds fully overlap at the final beam splitter. 
An extension of the standard Sagnac phase has been formulated and generalized for atoms evolving in an extended class of external potentials.
This can be used as a tool for designing improved potential configurations for Sagnac interferometry measurements.
\\

{\bf Acknowledgments:} PN gratefully acknowledges that this work was supported by the European Union's Seventh
Framework Programme (FP7-REGPOT-2012-2013-1) under grant agreement number 316165. 
This research has  also been supported by the EU-FET Grant FP7-ICT-601180 (MatterWave) and by
 FP7-PEOPLE-2012-ITN-317485 (QTea). We gratefully acknowledge useful discussions with the members of the MatterWave consortium. 

\newpage
\section{Bibliography}

\bibliography{bucketBibliography}

\newpage
\section{Appendix A: Determination of the potential} \label{sec:ApppendixAPotential}

\subsection{Coupling field calculation}

To determine the effective potential felt by the atoms, we use the rotating wave approximation (RWA) i.e.
we assume that only the polarized component of this field rotating perpendicularly to the quadrupole field
effectively interacts with the atoms. 
This perpendicular component is given by:
\begin{equation}
{\bf B}_\text{rf}^\perp({\bf r},t)={\bf B}_\text{rf}(t)-\frac{{\bf B}_\text{rf}(t). {\bf B}({\bf r},t)}{|{\bf B}({\bf r},t)|^2}{\bf B}({\bf r},t)~.
\end{equation}
This component rotates at the frequency $\omegarf$ with an elliptical polarization:
\begin{equation}\label{eq:perp}
\begin{aligned}
{\bf B}_\text{rf}^\perp({\bf r},t)&={\bf B}_1 \cos(\omegarf t)+ {\bf B}_2 \sin(\omegarf t)
\end{aligned}
\end{equation}
where 
\begin{eqnarray}
{\bf B}_1=
(B_+ +B_-)
\left(- \frac{B_\text{z} B_\text{x}}{|{\bf B}|^2},- \frac{B_\text{z} B_\text{y}}{|{\bf B}|^2},1- \frac{B_\text{z}^2}{|{\bf B}|^2}\right)~,
\\
{\bf B}_2=
(B_+ -B_-)
\left(1- \frac{B_\text{x}^2}{|{\bf B}|^2},- \frac{B_\text{y} B_\text{x}}{|{\bf B}|^2},- \frac{B_\text{z} B_\text{x}}{|{\bf B}|^2}\right)~.
\end{eqnarray}
It can be rewritten in the equivalent form:
\begin{equation}\label{eq:perp2}
\begin{aligned}
{\bf B}_\text{rf}^\perp({\bf r},t)
&={\bf B'}_1 \cos(\omegarf t +\chi)+ {\bf B'}_2 \sin(\omegarf t +\chi)
\end{aligned}
\end{equation}
with
a phase shift $\chi$  chosen such as to fulfill the orthogonality of the amplitudes 
${\bf B'}_1.{\bf B'}_2=0$. This last condition together with the comparison between the two last expressions Eq.(\ref{eq:perp}) and Eq.(\ref{eq:perp2}) allow to find
the following identities:
\begin{eqnarray}
\left(
\begin{array}{c}
{\bf B'}_1
\\
{\bf B'}_2
\end{array}
\right)
=\left(
\begin{array}{cc}
\cos \chi & -\sin \chi
\\
\sin \chi & \cos \chi
\end{array}
\right).
\left(
\begin{array}{c}
{\bf B}_1
\\
{\bf B}_2
\end{array}
\right)
\end{eqnarray}
and
\begin{eqnarray}
\tan (2\chi) = \frac{2{\bf B}_1.{\bf B}_2}{{\bf B}_2^2-{\bf B}_1^2}~.
\end{eqnarray}
Rewriting the amplitude of these field as $|{\bf B'}_1|=B'_+ +B'_-$
and $|{\bf B'_2|}=B'_+ -B'_-$, 
we can identify the new parameters as the effective Rabi coupling
$\hbar \Omega_{\pm}({\bf r},t)=g \mu_\text{B}{B'_\pm}$ and recover the equation (\ref{eq:coupling}).

A small distance expansion of the resulting effective potential Eq.(\ref{eq:VeffIntegral}) quadratic around the ring center ($r\sim R$ and $z\sim 0$) allows to identify the trap frequencies 
(\ref{eq:trappingFrequencies})  along the radial and 
z-direction respectively and the hyperfine dependent azimuthal components
in Eq.(\ref{eq:StateDependentRing}).

\section{Appendix B: Non adiabatic interferometry}\label{sec:AppndixBNonAdiabaticInterferometry}

In this appendix we discuss how to calculate the  phase evolution of  the atoms during their trajectory inside the ring. 
We  show that, if the atoms are at rest in the beginning and end of the interferometric sequence, then  the resulting phase depends only on the area enclosed and is essentially independent of the dynamic evolution of the atoms within the interferometer.
We generalize the Sagnac formula (\ref{eq:GeneralUncloseArea}) for the case that the trajectories do not fully enclose an area.
\subsection{The Sagnac phase in a travelling harmonic trap} \label{sec:UnitaryTransformation} 
We assume that the interactions between the atoms are negligible so that the atomic clouds in the two states $\ket{2,1}$ and $\ket{1,1}$ evolve according to the Schr\"odinger equation for the atoms in the inertial frame:
\begin{eqnarray}
i\hbar \partial_\text{t} \psi_\pm({\bf r},t)= 
-\frac{\hbar^2}{2m}\nabla_{\bf r}^2\psi_\pm({\bf r},t)+ V_\pm\left({\bf r}-{\bf r}_{0\pm}(t),t\right)\psi_\pm({\bf r},t) \,,
\end{eqnarray}
where the atoms are trapped in the moving harmonic potential $V_\pm({\bf r}-{\bf r}_{0\pm}(t),t)$ centered at the position ${\bf r}_{0\pm}(t)$ whose dynamics can be chosen arbitrarily. 
For the particular ring trap configuration, we use the dynamics
${\bf r}_{0\pm}(t)= \left\{R\cos\left[\pm\Omega t+ \phi_0(t)\right],\pm R\sin\left[\pm \Omega t + \phi_0(t)\right],0\right\}$ so that 
the potential equation (\ref{eq:fullPot}) is now rewritten in the inertial frame by taking into account of the external rotation angle $\Omega t$ that is added or subtracted to the angle of the trap minimum $\phi_0(t)$ in the laboratory frame.
In the trap center frame, the harmonic potential has the local form:
\begin{equation}\label{eq:hpot}
V_\pm({\bf r},t)=\frac{m}{2}{\bf r}.
\underline{\underline{{\mathbf \omega}}}^2_\pm(t).{\bf r}^T
\end{equation}
and depends on the trap position $\phi_0(t)$ through the tensor of the trap frequencies:
\begin{eqnarray}\label{om}
\underline{\underline{{\mathbf \omega}}}^2_\pm(t)=
\underline{\underline{{\mathbf O}}}(\Omega t \pm \phi_0(t)).
\left(\begin{array}{ccc}
\omega^2_r & 0 & 0 \\
0 & \omega^2_\phi & 0 \\
0 & 0 & \omega_\text{z}^2
\end{array} \right)
.
\underline{\underline{{\mathbf O}}}^T(\Omega t \pm \phi_0(t))
\end{eqnarray}
where we used the rotation matrix:
\begin{eqnarray}
\underline{\underline{{\mathbf O}}}(\phi)=
\left(\begin{array}{ccc}
\cos(\phi) & -\sin(\phi) &0 \\
\sin(\phi) & \cos(\phi) & 0\\
0 & 0 & 1
\end{array}\right)
\end{eqnarray}
This description can also be used for a ring shaped waveguide, with $\omega_\phi = 0$ and the trajectory of the trap center $\phi_0(t)$ being chosen such that it follows the free evolution of atom cloud.
We use the following transformations:
\begin{eqnarray}\label{psipm}
\psi_\pm({\bf r},t)= e^{i[{\bf k}_\pm(t).({\bf r}-{\bf r}_{\pm}(t))+\Phi_\pm(t)]}\psi_{0\pm}({\bf r}-{\bf r}_{\pm}(t),t)
\end{eqnarray}
to rewrite the Schr\"odinger equation in the moving frame of the position of the atom wave packet center ${\bf r}_{\pm}(t)$ as 
\begin{eqnarray}\label{moveframe}
i\hbar \partial_\text{t} \psi_{0\pm}({\bf r},t)=[-(\hbar^2/2m)\nabla_{\bf r}^2 + V_\pm({\bf r},t))]\psi_{0\pm}({\bf r},t)
\end{eqnarray}
This is achieved if this position ${\bf r}_{\pm}(t)$ obeys the dynamic equations:
\begin{eqnarray}\label{eq:pos}
{\bf \dot r}_{\pm}(t)&=&\hbar{\bf k}_{\pm}(t)/m
 \\ \label{mom}
\hbar{\bf \dot {\bf k}}_{\pm}(t)&=&-\frac{\partial V({\bf r}_{\pm}(t)-{\bf r}_{0\pm}(t),t)}{\partial{\bf r}_{\pm}(t)}
\\ \label{eq:action}
\Phi_{\pm}(t)&=&
\int_0^t \frac{dt'}{\hbar}L({\bf r}_{\pm}(t'),{\bf \dot r}_{\pm}(t'),t)
\nonumber \\ \label{eq:lag}
L_\pm({\bf r}_{\pm}(t),{\bf \dot r}_{\pm}(t),t)
&=&
\frac{1}{2}m{\bf \dot r}^2_{\pm}(t)-
V({\bf r}_{\pm}(t)-{\bf r}_{0\pm}(t),t).
\end{eqnarray}
This result is strictly only valid for a harmonic potential and is thus a good approximation
for atoms located close to the trap center
$|{\bf r}_{\pm}(t)-{\bf r}_{0\pm}(t)|\ll R$ (see Appendix \ref{anharm}).
As a consequence of the Kohn transformation that extends the translational invariance transformation concept to the case of the harmonic trap \cite{Kohn1961PR}, the center of mass position of the condensate center evolves according to the classical Lagrangian $L_\pm({\bf r}_{\pm}(t),{\bf \dot r}_{\pm}(t),t)$ \cite{Antoine2003PLA}. 
The transformation applied here to cartesian coordinates can be generalized to any time-dependent harmonic potential for any coordinates choice like the 
cylindrical coordinates developed in Appendix \ref{anharm} provided the resulting 
potential 
can be assumed quadratic in these coordinates. 

\subsection{Proof of the non adiabatic area theorem}

Let us 
suppose the rotation of angular velocity ${\bf \Omega}$ of the laboratory frame.
Then the laboratory coordinates ${\bf r}_L(t)$ are related to the inertial
coordinates through the relation ${\bf r}^T(t)=
\underline{\underline{{\mathbf O}}}({\bf \Omega}t) {\bf r}^T_L(t)$ where
we define the rotation matrix $\underline{\underline{{\mathbf O}}}({\bf \Omega}t)$ 
about the angular velocity at angle $|{\bf \Omega}t|$.
In this frame, the classical motion equation of an atom under an external potential
$V({\bf r}(t),t)$ becomes:
\begin{eqnarray}\label{momL}
m({\bf \ddot {\bf r}}_L(t)+ 2{\bf \Omega}\times { \dot {\bf r}}_L(t)
+{\bf \Omega}\times ({\bf \Omega}\times {\bf r}_L(t))&=&-\frac{\partial V({\bf r}_L(t),t)}{\partial{\bf r}_L(t)}
\end{eqnarray}
Making the perturbative development
${\bf r}_L(t)={\bf r}^{(0)}(t)+{\bf r}^{(1)}(t)+ \dots$
up to the first order in the angular velocity and applying it to Eq.(\ref{momL}), 
we determine the correction
to the action perturbatively. Suppose a classical motion between
time $t_A$ and $t_B$, we calculate successively:
\begin{equation}
\begin{aligned}
\int_{t_A}^{t_B}dt
&S({\bf r}_L(t),\dot{\bf r}_L(t),t)\\&=
\int_{t_A}^{t_B}dt\left[\frac{m}{2}
({\dot {\bf r}}_L(t)+ {\bf \Omega}\times {\bf r}_L(t))^2
-V({\bf r}_L(t),t)\right]\\
&=\int_{t_A}^{t_B}dt
S^{(0)}({\bf r}^{(0)}(t),\dot{\bf r}^{(0)}(t),t)
+2m{\bf \Omega}.\int_{{\bf r}^{(0)}(t_A)}^{{\bf r}^{(0)}(t_B)}{\bf r}^{(0)}\times d{\bf r}^{(0)}
+m {\bf \dot r}^{(0)}(t).{\bf r}^{(1)}(t)|_{t_A}^{t_B}
\end{aligned}
\end{equation}
Setting ${\bf r}^{(1)}(t_A)=0$, we find the result (\ref{eq:GeneralUncloseArea}).

\subsection{Anharmonic treatment}\label{anharm}

We generalize the results of section \ref{sec:UnitaryTransformation} to the case of an anharmonic equation.
In the case of strong radial confinement, the Schroedinger equation in the inertial frame is simplified into:
\begin{eqnarray}
i\hbar \partial_\text{t} \psi_\pm(\phi,t)=[-\frac{\hbar^2\partial_{\phi}^2}{2m R^2} - V_0(t) \cos(\phi \mp \phi_0(t)-\Omega t)]\psi_\pm(\phi,t)~.
\nonumber \\
\end{eqnarray}
Similarly to the harmonic treatment, the azimuthal part of the trap potential (\ref{eq:fullPot})
written in the inertial frame is centered about
$\pm\phi_0(t)+\Omega t$.
Thus according to the hyperfine
components, it accounts
for the external rotation angle $\Omega t$ that is added or substracted to
the trap minimum angle $\phi_0(t)$.
For the symmetric case where the atom components evolve in opposite directions,
the following transformation
\begin{eqnarray}
\psi_\pm(\phi,t)= e^{i [l_\pm (t) .(\phi - \phi_\pm(t)-\Omega t)+\Phi_\pm(t)]}
\psi_{0\pm}(\phi - \phi_\pm(t)-\Omega t)
\nonumber \\
\end{eqnarray}
rewrites the Schroedinger equation in the
moving frame of the atoms as:
\begin{eqnarray}\label{schanharm}
i\hbar \partial_\text{t} \psi_{0\pm}(\phi,t)&=&
\bigg \{-\frac{\hbar^2\partial_{\phi}^2}{2m R^2} + V_0(t)
[(1-\cos(\phi))
\cos(\phi_0(t)\mp \phi_\pm (t))
\nonumber \\&&-(\sin \phi -\phi) \sin(\phi_0(t) \mp \phi_\pm (t))]\bigg\}\psi_{0\pm} (\phi,t)
\end{eqnarray}
provided the following conditions are
fulfilled:
\begin{eqnarray}\label{posa}
{\dot \phi}_\pm(t)&=&\hbar{l}_{\pm}(t)/m R^2 -\Omega~,
 \\ \label{moma}
mR^2{\ddot \phi}_{\pm}(t)&=&-V_0(t)\sin(\phi_\pm(t) \mp \phi_0 (t))~,
\\
\Phi_{\pm}(t)&=&
\int_0^t \frac{dt'}{\hbar}L_\pm({\phi}_{\pm}(t),{\dot \phi}_{\pm}(t),t)~,
\nonumber \\ \label{laga}
L_\pm({\phi}_{\pm}(t),{\dot \phi}_{\pm}(t),t)
&=&\frac{\hbar^2({\dot \phi}_\pm(t)+ \Omega)^2}{2m R^2}
+V_0(t)\cos(\phi_0(t)\mp \phi_\pm (t))~.
\nonumber \\
\end{eqnarray}
With the boundary conditions ${\phi}_{\pm}(0)= {\dot \phi}_{\pm}(0)=0$, we
recover the formulae (\ref{eq:IncompleteSagnacPhase}) whatever the dynamics involved.
Let us note that $l_\pm(t)$ should strictly speaking be an integer since the wave function should be periodic in $\phi$.
However as long as the radius is much larger than the spatial extension of the wave function ($R \gg a_\text{ho}$), it can be extended to the real value domain.
Under the condition of a localized wave function, the equation (\ref{schanharm}) is  simplified into an harmonic oscillator:
\begin{eqnarray}\label{schanharm1}
i\hbar \partial_\text{t} \psi_{0\pm}(\phi,t)=\left\{-\frac{\hbar^2\partial_{\phi}^2}{2m R^2} + V_0(t)
\cos[\phi_0(t)\mp \phi_\pm (t)]\frac{\phi^2}{2}\right\}\psi_{0\pm} (\phi,t)\,
\end{eqnarray}
where the effective trap frequency is time dependent: 
\begin{equation}
 \nonumber \omega^\text{eff}_{\phi\pm}(t)= \sqrt{V_0(t)\cos(\phi_0(t)\mp \phi_\pm (t))/mR^2}\,.
\end{equation}
In order to avoid further phase accumulation, strictly speaking, one has to ensure the adiabaticity of the evolution of the wave function. 
The main limitation is the transition to the excited state in the trap which is negligible 
when the transit time $T$ is much longer than the time variation of the trap frequency i.e.~$T {\dot \omega^\text{eff}_{\phi\pm}(t)}/\omega^\text{eff}_{\phi\pm}(t) \ll 1$. 
All these conditions fulfilled, we recover using (\ref{laga}) the formula (\ref{eq:IncompleteSagnacPhase}) whatever the dynamics involved.
In practice, these conditions are only fulfilled in the bucket case where

using Eq.(\ref{eq:phit}) we find $\omega_\phi T \gg 1$.
Nevertheless, this additional phase accumulation is not important as it cancels out in the phase difference if we restrict the calculation in the linear contribution in $\Omega$.

\section{Appendix C: Contrast}\label{sec:AppendixCContrast}

The atom profile after recombination at the second 
$\pi/2$ pulse at time $t$ is altered when the wave packets coming 
from the two arms are not synchronized.
The contrast resulting from these imperfections can be determined quantitatively from the atom density obtained for the two states after interferometry:
\begin{eqnarray}\label{Prt}
P_\pm({\bf r},T)&=&\frac{1}{4}
|\psi_+({\bf r},t)\pm \psi_-({\bf r},t)|^2~.
\end{eqnarray}
Because of the high confinement in along the radial direction, we neglect the effect of the internal rotation inside the trap during the interferometry process.
Using Eq.(\ref{psipm}) together with the ground state wave function:
\begin{eqnarray}\label{gs}
\psi_{0\pm}({\bf r})
=\left(\frac{m\omega_{ho}}{\pi \hbar}\right)^{3/4}
\exp[-\frac{m}{2\hbar}{\bf r}.
\underline{\underline{{\mathbf \omega}}}_\pm (t).{\bf r}^T]
\end{eqnarray}
and $\omega_{ho}=(\omega_\text{r} \omega_\text{z} \omega_\phi)^{1/3}$,
we obtain after integration the probability of the two hyperfine 
states after interferometry:
\begin{eqnarray}
\nonumber P_\pm(T)&=&\int d^3{\bf r} P_\pm({\bf r},T)
\nonumber \\&=& \frac{1}{2}\bigg(1 \pm {\cal V}(T)\cos\left[\Phi_\text{S}(T)-\frac{{\bf k}_{+}(T)+{\bf k}_{-}(T)}{2}.\delta {\bf r}(T)\right]
\bigg)
\end{eqnarray}
where we define the visibility:
\begin{eqnarray}
{\cal V}(T)=
\exp\left[-\frac{m^2 \delta {\bf r}(T).
\underline{\underline{{\mathbf \omega}}}_\pm(t).\delta {\bf r}^T(T)+
\delta {\bf k}(T).
\underline{\underline{{\mathbf \omega}}}^{-1}_\pm(t).\delta {\bf k}^T(T)
}{4\hbar m}\right]
\end{eqnarray}
depending on the mass center position 
$\delta {\bf r}(t)={\bf r}_{+}(T)-{\bf r}_{-}(T)$ and the momentum difference 
$\delta{\bf k}(t)= {\bf k}_{+}(T)-{\bf k}_{-}(T)$.
The contrast of the interferometer will be reduced, if the two wave packets originating from the two arms of the interferometer have a finite velocity relative to each other or 
if they do not overlap perfectly at the final beam splitter.
In the case of a tight confinement within the ring, the probability simplifies into:
\begin{eqnarray}\label{eq:contrast}
P_\pm(T)&=&\frac{1}{2} \pm \frac{{\cal V}(T)}{2}
\cos\left[\Phi_\text{S}(T)-\frac{mR^2}{\hbar}\frac{{\dot \phi}_{+}(T)+{\dot \phi}_{-}(T)}{2}(\delta {\phi}(T)-4\pi)\right]
\end{eqnarray}
\vspace{3mm}
where
\begin{eqnarray}\label{eq:visibility}
{\cal V}(T)=
\exp\left[-\frac{m R^2[\omega_\phi^2 (\delta {\phi}(T)-4\pi)^2+\delta{\dot \phi}^2(T)]}{4\hbar\omega_\phi}\right]~.
\end{eqnarray}
Using the phase dynamics Eq.(\ref{eq:phit}) for the bucket case, the visibility has the 
explicit expression:
\begin{eqnarray}\label{eq:visibilitybucket}
\ln {\cal V}(T)=-\frac{m R^2\omega_\phi}{\hbar}
\left(\frac{4\pi\sin(\omega_\phi T/2)}{\omega_\phi T}\right)^2~.
\end{eqnarray}
Thus unless we are in the resonance region $T=2\pi n/\omega_\phi$ 
or in the large limit $T\omega_\phi \rightarrow \infty$
the visibility is less than one.

\end{document}